%% file: siribase_paper_sigmod.tex
\documentclass[sigconf]{acmart}
\AtBeginDocument{%
  \providecommand\BibTeX{{%
    \normalfont B\kern-0.5em{\scshape i\kern-0.25em b}\kern-0.8em\TeX}}}

\copyrightyear{2022} 
\acmYear{2022} 
\setcopyright{acmlicensed}\acmConference[SIGMOD '22]{Proceedings of the 2022 International Conference on Management of Data}{June 12--17, 2022}{Philadelphia, PA, USA}
\acmBooktitle{Proceedings of the 2022 International Conference on Management of Data (SIGMOD '22), June 12--17, 2022, Philadelphia, PA, USA}
\acmPrice{15.00}
\acmDOI{10.1145/3514221.3526049}
\acmISBN{978-1-4503-9249-5/22/06}

\usepackage{xspace}

\definecolor{afblue}{rgb}{0.36, 0.54, 0.66}

\def\akp{Saga\xspace}
\def\akg{KG\xspace}

\keywords{knowledge graphs, knowledge graph construction, entity resolution, entity linking}


\begin{document}
\fancyhead{}
\title{\akp: A Platform for Continuous Construction and Serving of Knowledge At Scale}


\author{Ihab F. Ilyas, Theodoros Rekatsinas, Vishnu Konda \\Jeffrey Pound, Xiaoguang Qi, Mohamed Soliman}
\affiliation{
	\institution{Apple}
	\city{}
	\country{}
	}

\renewcommand{\shortauthors}{Ilyas, Rekatsinas et al.}

\begin{abstract}
We introduce \akp, a next-generation knowledge construction and serving platform for powering knowledge-based applications at industrial scale. 
\akp follows a hybrid batch-incremental design to continuously integrate billions of facts about real-world entities and construct a central knowledge graph that supports multiple production use cases with diverse requirements around data freshness, accuracy, and availability.
In this paper, we discuss the unique challenges associated with knowledge graph construction at industrial scale, and review the main components of \akp and how they address these challenges.
Finally, we share lessons-learned from a wide array of production use cases powered by \akp.
\end{abstract}

\maketitle

\input{sections_sigmod/intro.tex}
\input{sections_sigmod/construction.tex}
\input{sections_sigmod/compute.tex}

\input{sections_sigmod/livegraph.tex}
\input{sections_sigmod/services.tex}

\input{sections_sigmod/usecases.tex}
\input{sections_sigmod/related.tex}

\input{sections_sigmod/conclusion.tex}

\subsection*{Acknowledgements}
This work was made possible by Omar Attia, Ryan Clancy, Mina Farid, Ahmed Fakhry, Dylan Fu, Ankur Goswami, Nahush Kulkarni, William Ma, Ali Mousavi, Victor Suthichai, Aadithya Udupa, 
Varun Notibala, Niharika Bollapragada, Rifat Ahsan, Ramesh Balaji, Mukund Sharma, Eric Choi, Abhishek Tondehal, Jennifer Cooper, Hans Wang and many others. We thank many teams at Apple for support and feedback.

\bibliographystyle{ACM-Reference-Format}
\bibliography{siribase_paper.bib}

\end{document}

%% file: sections_sigmod/intro.tex

\section{Introduction}\label{sec:intro}

%
%
%

Accurate and up-to-date knowledge about real-world entities is needed in many applications. Search and assistant services require open-domain knowledge to power question answering. Other applications need rich entity data to render entity-centric experiences. Many internal applications in machine learning need
training data sets with information on entities and their relationships. All of these applications require a broad range of knowledge
that is accurate and continuously updated with facts about entities.
Constructing a central knowledge graph (KG) that can serve these needs is a challenging problem, and developing a KG construction and serving solution that can be shared across applications has obvious benefits.
This paper describes our effort in building a next-generation knowledge platform for continuously integrating billions of facts about real-world entities and powering experiences across a variety of production use cases.


Knowledge can be represented as a graph with edges encoding \emph{facts} amongst \emph{entities} (nodes)~\cite{industry_kgs}.
Information about entities is obtained by integrating data from multiple structured databases and data records that are extracted from unstructured data~\cite{deepdive_sigmod_record}.
The process of cleaning, integrating, and fusing this data into an accurate and canonical representation for each entity is referred to as \emph{knowledge graph construction}~\cite{kg_book}.
Continuous construction and serving of knowledge plays a critical role as access to up-to-date and trustworthy information is key to user engagement.
The entries of data sources used to construct the \akg are continuously changing: new entities can appear, entities might be deleted, and facts about existing entities can change at different frequencies.
Moreover, the set of input sources can be dynamic. 
Changes to licensing agreements or privacy and trustworthiness requirements can affect the set of admissible data sources during KG construction.
Such data feeds impose unique requirements and challenges that a knowledge platform needs to handle:
 \begin{figure}[t]
        \centering
      \includegraphics[width=0.8\columnwidth]{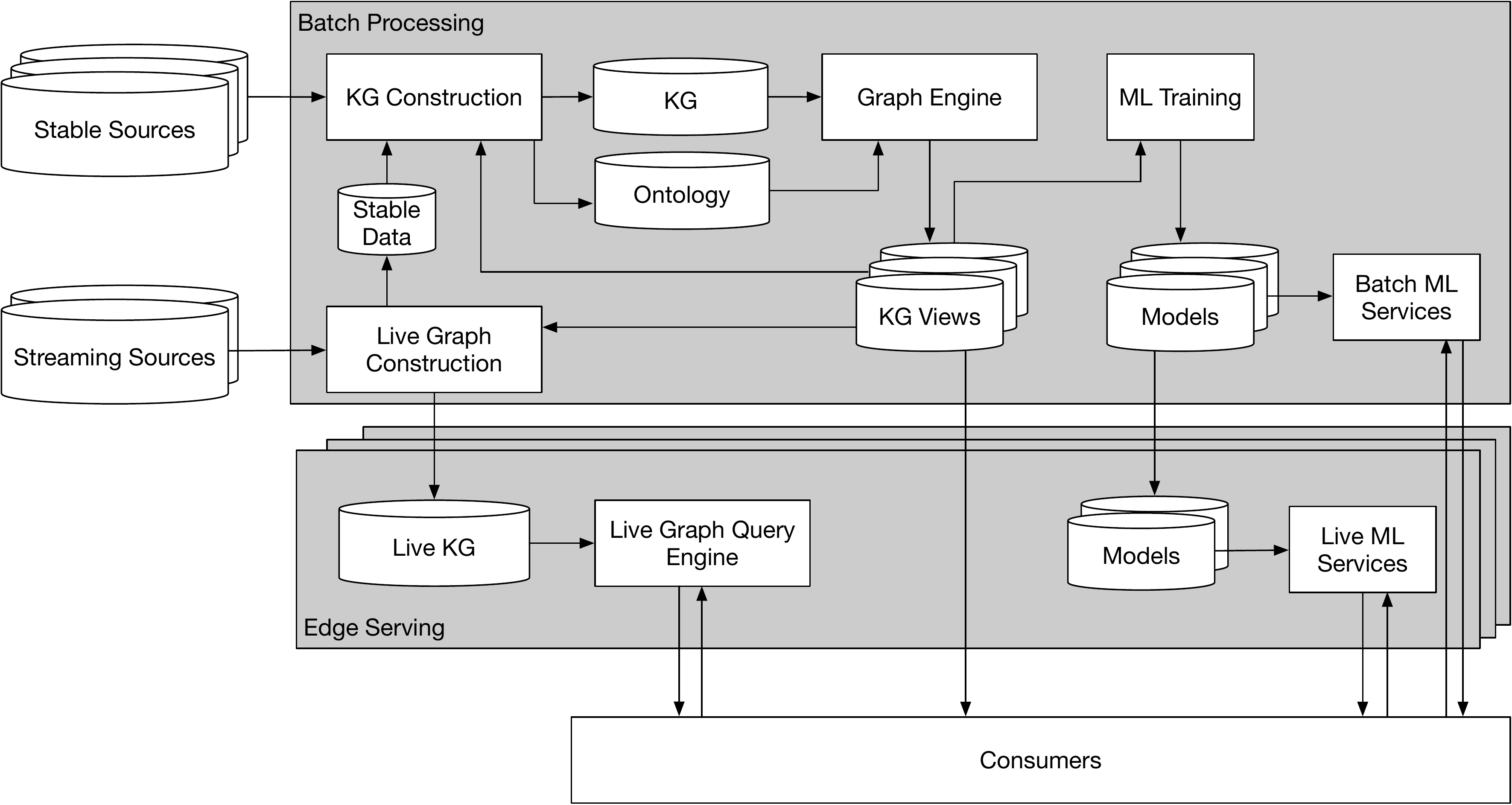}
      \caption{Overview of the \akp knowledge platform.}
    \label{fig:arch}
\end{figure}
\begin{enumerate}
	\item \emph{Hybrid batch and stream construction}: Knowledge construction requires operating on data sources over heterogeneous domains. The update rates and freshness requirements can differ across sources. Updates from streaming sources with game scores need to be reflected in the KG within seconds but sources that focus on verticals such as songs can provide batch updates with millions of entries on a daily basis. Any platform for constructing and serving knowledge has to provide support for batch and stream processing with service-level agreements (SLAs) around data freshness, end-to-end latency, and availability.
	\item \emph{Provenance management}: Attribution, reliability control, and license management are key ingredients in a knowledge platform. Transparency is critical for serving data to production use cases (e.g., intelligent assistants) that surface knowledge information; all facts in the KG are required to carry data provenance annotations for data governance purposes. Any knowledge platform needs to adhere to non-destructive data integration procedures that enable surfacing the provenance of individual facts, serving KG views that conform to licensing agreements, and enforcing on-demand data deletion.
	\item \emph{Targeted fact curation}: To ensure an engaging user experience for entity-rich services, the information in the KG needs to be correct and up-to-date. Accuracy, coverage, and freshness of the served knowledge are key requirements. To meet these requirements, processes that enable continuous and incremental acquisition, integration, and verification of new facts in a targeted and on-demand manner are critical features in a knowledge platform.
	\item \emph{Knowledge graph views and computed knowledge artifacts}: Many production use cases rely on data artifacts computed over the KG (such as computed entity importance measures) to provide entity-rich experiences to users. It is critical that any knowledge platform supports multiple data consumers and allows them to register and continuously consume custom views of the KG. This functionality requires  a graph query engine that supports rich view definitions and materialization policies while ensuring compliance to privacy policies for different registered views.
	\item \emph{Self-serve data onboarding}: Low-effort onboarding of new data sources is important to ensure consistent growth of the KG. Any knowledge platform needs to provide APIs that allow domain teams to develop and deploy data pipelines that will allow continuous integration of their data in the KG. Self-serve-centric and modular APIs are required to ensure ease-of-use and extensibility.
	\item \emph{Run-time indexes and APIs}: The KG is the backbone of entity-centric Question Answering and entity-centric experiences (such as Entity Cards). Meeting the SLAs imposed by those user-facing services requires constructing knowledge indexes that can serve structured queries over the KG with strict latency requirements and can also be updated in real time to reflect the most recent information about entities.
	\item \emph{Semantic annotations service}: The KG offers a controlled vocabulary that can be used to enrich data in production cases with entity-centric information. A \emph{semantic annotation service} that can tag data from different organizations and verticals with concepts and entities in the KG is a fundamental component of any knowledge platform. This service must operate on diverse inputs, e.g., structured and unstructured data, and provide accurate annotations for both head (i.e., popular) and tail (i.e., less popular) entities and concepts.
\end{enumerate}

\begin{table}[t]
\scriptsize
  \begin{tabular}{llllllll}
    \toprule
    subj & predicate & r\_id & r\_predicate & obj & locale & sources & trust \\
    \midrule
   	e1 & name &  &  & J. Smith  & en & [src1, src2] & [0.9, 0.8] \\
      	e1 & educated\_at & r1 & school  & UW & en & [src2] & [0.8] \\
	e1 & educated\_at & r1 & degree  & PhD & en & [src2] & [0.8] \\
	e1 & educated\_at & r1 & year  & 2005 & en & [src2] & [0.8] \\
    \bottomrule
  \end{tabular}
    \caption{Extended triples representation of the KG in Figure~\ref{fig:kg}. Symbols r\_id and r\_predicate are abbreviations of relationship id and relationship predicate.}
  \label{tab:ext}
\end{table}

This paper introduces \akp, a next-generation knowledge construction and serving platform for powering knowledge-based applications at industrial scale. The paper describes the system considerations and design decisions we followed to build \akp and reviews deployments that power industrial use cases.
The paper is organized by technical theme and covers key parts of the architecture of \akp (see Figure~\ref{fig:arch}). 

%% file: sections_sigmod/construction.tex

\section{Knowledge Graph Construction}\label{sec:construction}
%
%
%

Knowledge Graph Construction is the process of integrating multiple diverse data sources into a standardized repository of linked entities and concepts~\cite{kg_book}. In our case, data sources range from open-domain and general knowledge sources such as Wikipedia and Wikidata to specialized sources that provide data about music domains, media products, sports, celebrities, nutritional substances and many more. The KG provides a succinct and integrated representation of all \emph{entities} that appear in these sources, including the \emph{predicates} (attributes) related to each entity and the \emph{relationships} among these entities. This representation follows an in-house open-domain ontology. The ontology is designed such that it also enables a data model that allows for optimized processing of large volumes of graph-structured data feeds. Next, we review the data model adopted by the \akg, we introduce \emph{data source ingestion} and \emph{knowledge construction}, two core modules that facilitate building and maintaining the KG. Finally, we discuss how \akp supports scalable and incremental knowledge construction.


\subsection{Data Model}\label{sec:data_model}
To represent the KG, \akp follows the RDF data model format with \texttt{<subject,predicate,object>} triples~\cite{rdf} . Each entity is represented as a set of triples. Each triple states a \emph{fact} such as the name of a person, or the capital of a country. Relationships are represented by linking different entities: \texttt{object} can either be a literal value or a reference to another entity. This structure defines a \emph{directed graph}, where predicates represent edges and subjects or objects represent nodes.

 \begin{figure}[t]
        \centering
      \includegraphics[width=0.6\columnwidth]{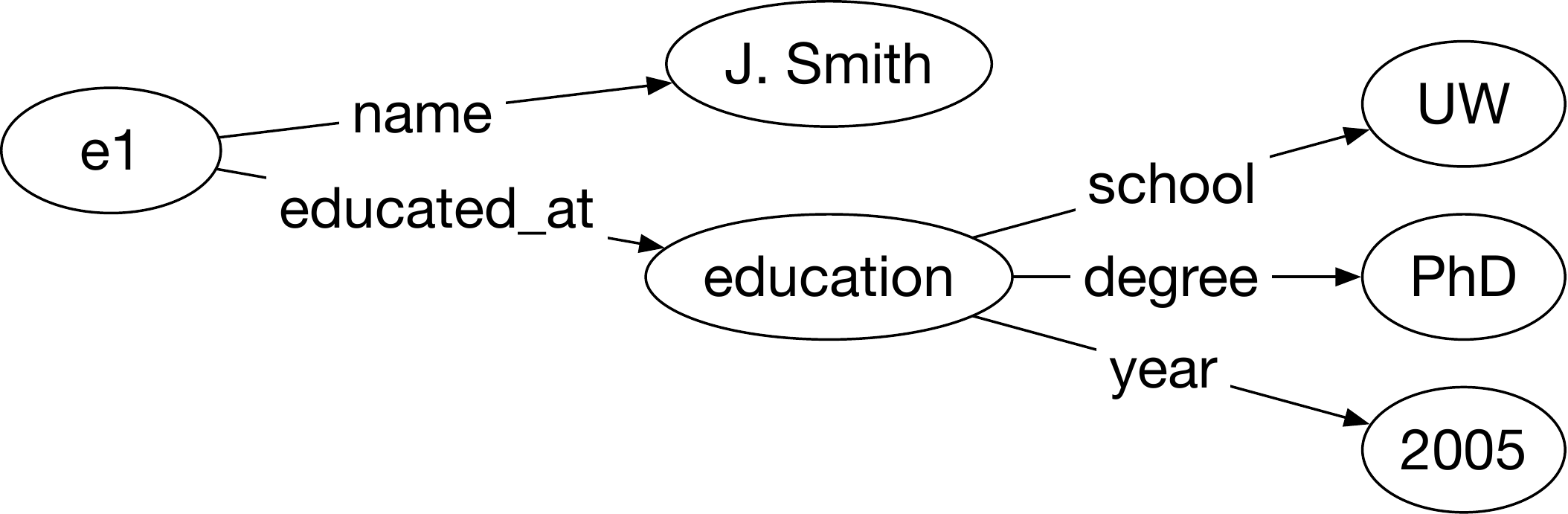}
      \caption{An example KG for the facts in Table~\ref{tab:ext}.}
    \label{fig:kg}
\end{figure}

Consider the example KG in~\autoref{fig:kg} about persons and their education. Subject $e_1$ is has a $name$ predicate that points to a literal object `J. Smith'. Relationships among entities are often composite with additional structure. To illustrate, consider for example the predicate $educated\_at$ that associates $e_1$ to a composite $education$ object, which in turn has  $school$, $degree$ and $year$ predicates.

To facilitate retrieval of properties from linked entities, the triple representation is extended to capture one-hop relationships among  entities. For example, the predicate  $educated\_at$  in~\autoref{fig:kg} is represented using a set of triples to capture composite attributes such has $educated\_at.school$ as part of the facts describing the main entity $e_1$. We call this representation {\em extended triples}, as shown in~\autoref{tab:ext}. Extended triples provide a flat relational model of the KG. This data model allows easy retrieval of the frequently used one-hop relationship data without performing an expensive self-join or graph traversal operation. The extended triples format is a variation of the JSON-LD format~\cite{json_ld}, a lightweight Linked Data format adopted by industry-scale KGs for efficient querying~\cite{extended_triples}.

Finally, we augment the extended triple format with metadata fields that track the \emph{provenance (sources)}, \emph{locale}, and \emph{trustworthiness} for each fact. To track provenance, we associate each record with an array of references to input data sources. This array is always updated to track the integration of records from multiple sources to construct a single record in the KG. This approach allows us to attribute every fact to its data sources and provides a mechanism to ensure compliance with the source license agreements. Locale-specific metadata are associated with literals and string objects in the KG. This information is important for storing multi-lingual knowledge. Finally, each KG record is associated with a trustworthiness score array, corresponding to record sources. These scores are used to obtain an aggregated confidence score on the correctness of each record. Prior works have also considered associating every fact in a KG with a correctness score~\cite{knowledge_trust}. Confidence scores provide a probabilistic representation of knowledge, which allows for accuracy SLA's and drives fact auditing decisions.

\subsection{Data Source Ingestion}\label{sec:source_ingestion}
This Data Source Ingestion module of \akp is composed of a set of pluggable and configurable adapters that implement the steps needed to ingest and onboard data from a given provider into the KG. Multiple challenges need to be addressed in this regard:

\begin{itemize}
\item Support different data formats (e.g., Parquet files in HDFS, CSV, JSON etc.) by providing a repository of data importers that support different formats.
\item Align the data about entities from different data sources to the ontology of the KG by providing a configurable  interface to specify \emph{ontology alignment} constructs, as well as scalable processing of these constructs.  
\item Export the aligned source data for consumption by the KG construction pipeline. Data needs to be exported as extended triples for efficient onboarding to the KG.
\end{itemize}

 \begin{figure}[t]
        \centering
      \includegraphics[width=1\columnwidth]{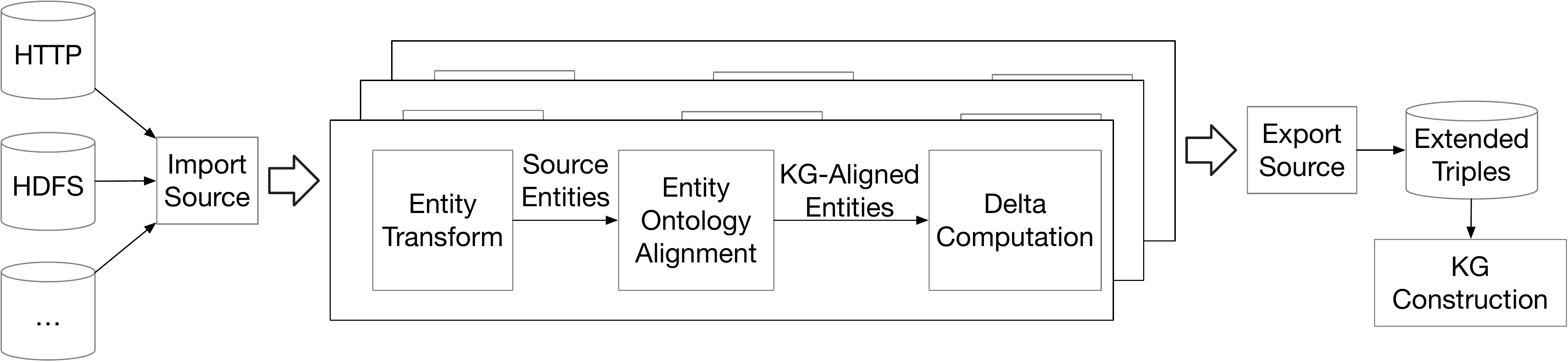}
      \caption{The source ingestion module of \akp.}
    \label{fig:ingest}
\end{figure}

\autoref{fig:ingest} illustrates the source ingestion pipeline, which ingests one or more entity payloads from upstream data provider and ensures data compliance with the KG data format and ontology. Each ingestion pipeline has multiple stages:
\begin{itemize}
\item \emph{Import}: read upstream data in their raw format into rows; each imported row may contain a single or multiple entities.
\item \emph{Entity Transform}: produce entity-centric views from the imported source data.
    Each row in the output of the transformation phase captures \emph{one entity}, and its columns capture entity predicates expressed in the source namespace.
\item \emph{Ontology Alignment}: populate a target schema that follows the KG ontology.
 In this stage, source entities are consumed as input and corresponding entities are produced as output. The {\em predicates} of output entities follow the KG ontology, while the {\em subject} and {\em object} fields remain in the original data source namespace; they are later linked to KG entity identifiers during knowledge construction. Entity type specification is also part of this step. This alignment is manually defined and controlled via configuration files.
\item \emph{Delta Computation}: detect changes with respect to the previously consumed snapshots of source data. This step crucial to determine what has changed in the upstream data, and subsequently minimize the volume of data consumed by knowledge construction. Change detection is performed eagerly: when an upstream provider publishes a new data version, the difference with respect to the data already consumed by \akp is computed and materialized to be picked up by knowledge construction.
\item \emph{Export}: generate extended triples in the KG-ontology schema to be consumed by knowledge construction. 
\end{itemize}    

Extensibility is key for quick and scalable onboarding of new data sources. To build a new source ingestion pipeline, engineers only need to provide implementation of the following interfaces:

\paragraph{Data Source Importer} This component reads upstream data artifacts and converts them  into a standard row-based dataset format. This component is responsible for normalizing the heterogeneity of upstream data for the rest of the pipeline by reading source data artifacts into a unified representation. For example, we may need to combine raw artist information and artist popularity datasets to get complete artist entities. \akp provides importer templates that can be altered to develop custom source ingestion pipelines.

\paragraph{Data Transformer} This component consumes a uniform data representation from importers and produces an entity-centric view of the upstream data source. Each entity is represented as a multicolumn row and columns are used to represent source predicates. The data transformer allows joining multiple data artifacts together to obtain a comprehensive description of a source entity. The transformer does not add any new predicates but allows implementing data integrity and sanity checks:
\begin{itemize}
\item Entity IDs are unique across all entities produced.
\item Each entity must have an ID predicate. This constraint is crucial to uniquely identify source data entries after they are added to the KG and key to incremental KG construction.
\item Predicates must be non-empty.
\item The predicates in the source schema are present in the produced entity (even if they are null/empty).
\item Predicate name must be unique in the source entity.
\end{itemize} 

\paragraph{Predicate Generation Functions (PGFs)} These lightweight methods are used to align the source entity data with the target schema and format of the KG. The concept of PGFs is related to that of \emph{tuple-generating dependencies}~\cite{integration_book}. For ease of use, \akp uses a config-driven development paradigm. Users specify both the source predicates and target predicates from the KG ontology in the configuration. Then, PGFs based on this specification are used to populate the target schema from the source data. These methods define the alignment of source predicates to KG predicates. To illustrate, consider a movies data source. When movie entities are ingested, they may be described in a source-specific schema and namespace. To standardize such input against the KG, alignment of ontologies needs to be done. A predicate in the source entity could be mapped to a predicate with a different name in the target ontology (e.g., \texttt{category} is mapped to \texttt{genre}). Similarly, a group of predicates may need to be combined to produce a target predicate (e.g., \texttt{<title, sequel\_number>} is mapped to \texttt{full\_title}).  

\subsection{Knowledge Construction}\label{sec:knowledge_construction}
Given the ontology-aligned source data, we need to integrate the extended triples from the input sources with the KG. Recall that at this point the subjects and objects are still not standardized. The goal of knowledge construction is to standardize the subjects and objects to refer to appropriate entities in the KG. We need to address the next technical problems for high-accuracy integration:
\begin{itemize}
\item \emph{In-source Deduplication}: Input sources can have duplicate entity records, hence, we need to detect duplicates within their records. Moreover, we need to store the necessary metadata that will allow us to consolidate these duplicates in later steps of knowledge construction. 
\item \emph{Subject Linking}: Given a source entity, identify an existing KG entity that corresponds to the same real-world entity intended by the source. If such an entity is found, the source entity is {\em linked} by getting assigned a KG entity identifier.
\item \emph{Object Resolution (OBR)}: In many cases, a string literal, e.g., person name, is given as the {\em object} field of a triple. To ensure accurate integration, we need to either resolve this string literal into an existing KG entity or create a new entity. This step normalizes the cross-references among KG entities.
\item \emph{Fusion}: Given a linked source entity and the meta-data from in-source duplication, we extend the KG entity repository with new and/or updated facts based on the source payload. It is important that we consolidate facts across duplicate and linked entities to obtain a consistent KG.
\end{itemize} 

 \begin{figure}[t]
        \centering
      \includegraphics[width=0.8\columnwidth]{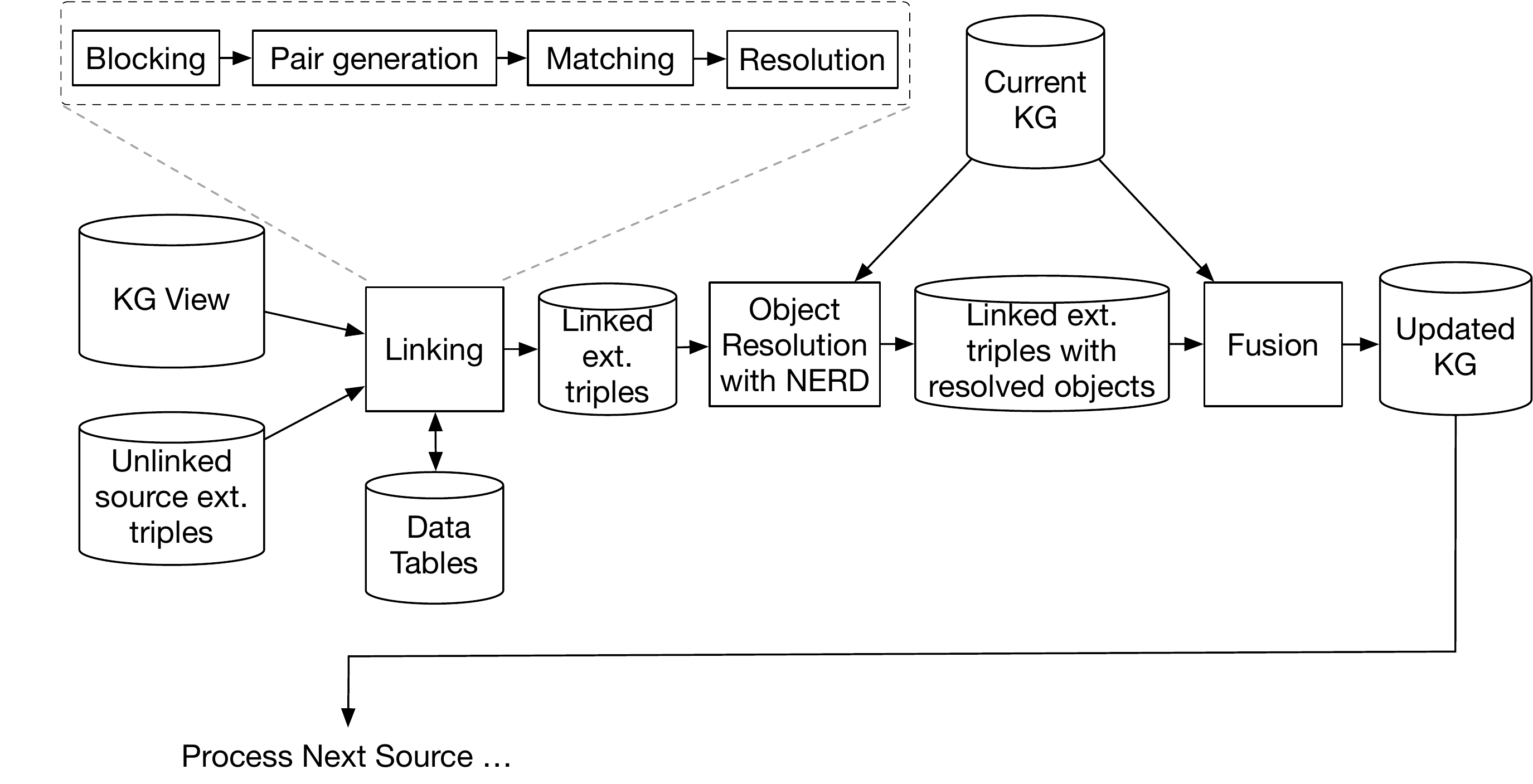}
      \caption{A pipeline for knowledge construction.}
    \label{fig:construction}
\end{figure}

The architecture of the knowledge construction pipeline of \akp is shown by~\autoref{fig:construction}. We next describe the pipeline stages.

\paragraph{Linking} This stage addresses the technical problems of in-source deduplication and subject linking. Both problems correspond to instances of \emph{Record Linkage} where different instances of the same real world entities need to be identified~\cite{koudas2006record, getoor2012entity, stonebraker2013data, integration_book}. Linking is performed by conducting the following steps:
\begin{enumerate}
\item  Input data is grouped by entity type. For each entity type, e.g., movies, we extract a subgraph from the current KG containing relevant entities. This step reduces the scope of entity linking to a smaller target dataset. We call this subgraph a \emph{KG view} (see Section~\ref{sec:views}).
\item We combine the source entity payload (which may include duplicates) with the KG view into one combined payload over which we perform record linking.
\item We apply \emph{blocking} on the combined payload~\cite{whang2009entity,papadakis2020blocking, getoor2012entity, steorts2014comparison}. During blocking, entities are distributed across different buckets by applying lightweight functions to group the entities that are likely to be linked together, e.g., a blocking function may group all movies with high overlap of their title $q$-grams into the same bucket. The goal is to partition data into smaller groups of potentially highly similar entities within each group and hence reduce the inherent quadratic complexity of the record linking problem. 
\item Given the blocking step output, we generate pairs of entities that belong to the same block. Then, a \emph{matching model}~\cite{wang2011entity} computes a similarity score for each pair of entities. Matching models are domain-specific and focus on specific entity types controlled by the ontology. The matching model emits a calibrated probability that can be used to determine if a pair of entities corresponds to a true match or not. The platform allows for both machine learning-based~\cite{mudgal2018deep, doan2020magellan} and rule-based matching models~\cite{singh2017generating, elmagarmid2014nadeef}. \akp offers a wide array of both deterministic and machine learning-driven similarity functions that can be used to obtain features for these matching models. We discuss learned similarity functions in more detail in Section~\ref{sec:ml}.
\item The last step in Linking is that of \emph{resolution}. Given the probability of similarity for all relevant entity pairs, we find entity clusters that correspond to the same real-world entity~\cite{bhattacharya2007collective, collective_entity_matching, saeedi2017comparative}. To ensure scalability, we use the calibrated similarity probabilities to identify high-confidence matches and high-confidence non-matches and construct a \emph{linkage graph} where nodes correspond to entities and edges between nodes are annotated as \emph{positive} (+1) or \emph{negative} (-1). We use a correlation clustering algorithm~\cite{parallel_correlation_clustering} over this graph to identify entity clusters. During resolution, we require that each cluster contains at most one \emph{graph entity}. For all source entities in a cluster, we assign the identifier of the graph entity. If no graph entity exists in the cluster, we create a new KG entity and assign the identifier of the new entity to all source entities. Additional \emph{same\_as} facts that record the links between source entities and KG entities are  maintained to provide full provenance of the linking process. 
\end{enumerate} 

The previous steps need to be repeated when onboarding data from different  entity types, e.g., artist, song, album, etc, since each entity type can have domain-specific logic for blocking and matching. To scale the computation, processing within each block can be parallelized and the generation of linking artifacts happen incrementally as more blocks get processed.

\paragraph{Object Resolution} Mapping string literals or id values in the {\em object} field into KG entity identifier is the goal of the Object Resolution (OBR) step~\cite{grishman-sundheim-1996-message}.
A machine learning framework for Named Entity Recognition and Disambiguation (NERD) is used to map entity names based on the context in which they appear, to graph entity identifiers. We describe our NERD architecture in Section~\ref{sec:ml}.

\paragraph{Fusion} Given a collection of linked source entities, fusion addresses the problem of merging the source payload with the KG to take it into a new consistent state~\cite{dong2009data,dong2013big,li2016survey}. For simple facts that are given directly by a predicate in the source triples, e.g., birthdate, these can be fused by performing an outer join with the KG triples. This will either update the source provenance of facts in the graph, or add a new fact if it does not exist. For composite facts given by a combination of {\em predicate/relationship\_predicate} (cf. Figure~\ref{fig:kg}), fusion needs to be more elaborate in order to judge if the source relationship node can be merged with an existing KG relationship node, or it needs to be added as a completely new relationship node. This operation is done by estimating the similarity of facts in relationship nodes in both the source entity payload and the KG entity payload. A pair of relationship nodes with sufficient intersection in their underlying facts is deemed similar and can be merged together. All other relationship nodes in the source payload are added as new relationship nodes to the KG. During fusion, we use standard methods of truth discovery and source reliability methods~\cite{slimfast, luna_fusion, knowledge_trust, heidari2020record} to estimate the probability of correctness for each consolidated fact. These algorithms reason about the agreement and disagreement across sources and also take into account ontological constraints. The associated probability of correctness is stored as metadata in the KG and used by downstream tasks such as targeted fact curation (see Section~\ref{sec:usecases}).

\subsection{Scaling Knowledge Graph Construction}\label{sec:scaling_kgc}
The design of \akp exploits parallelism opportunities to significantly reduce the end-to-end construction time. To cope with the nature of continuous changes  in the underlying data sources (e.g., a new movie was released, or a song popularity got updated), source data preparation needs to be offloaded to the source ingestion platform. The disparate and parallel nature of ingestion pipelines of different sources provides an opportunity for scalability, where all source-specific processing is conducted in parallel to prepare payloads for consumption by the KG construction pipeline. 

In this regard, two key functionalities of the source ingestion platform are (i) generation of extended triples in the KG namespace, and (ii) eager computation of source deltas with respect to the latest snapshot consumed by the KG, following an incremental knowledge construction paradigm~\cite{gruenheid2014incremental, yan2020entity, stonebraker2013data}. A partitioned dump of source data is eagerly generated as follows. Let $t_0$ be the last timestamp a source has been consumed by the KG, and $t_n$ is the current timestamp, source ingestion pipeline splits source entities into three partitions:
\begin{itemize}
\item \emph{Added}: all source entities that exist at $t_n$ but not at $t_0$ 
\item \emph{Deleted}: all source entities that exist at $t_0$ but not at $t_n$ 
\item \emph{Updated}: all source entities that exist at both $t_0$ and $t_n$ and are modified at $t_n$.
\end{itemize}  

              
%


In addition, a separate full dump of triples capturing volatile predicates (e.g., entity popularity) of all source entities is produced. Changes in these predicates are not reflected in the above dumps. This is important to factor-out update churns (e.g., movie popularity might be updated very frequently) from delta payloads.


Knowledge construction is designed as a continuously running delta-based framework; it always operates by consuming source diffs. When a completely new source needs to be consumed, it is captured as a source with a full \emph{Added} payload and empty \emph{Deleted} and \emph{Updated} payloads. The end result of construction pipeline is an updated KG that reflects the latest source data changes. 

The linking pipelines of different data sources are run in parallel to allow for scalable construction. The main functionality needed to allow this mode of operation  are the following:

\begin{itemize}
\item \emph{Lightweight Ingestion}: Ingestion of changed source data into construction pipeline is largely simplified. For example, the extended triples from each source already provide the needed triplication of composite relationship nodes, and so self joins on ingested source data to compute one hops is avoided. 

\item \emph{Source-based Enrichment}:  Linking may require joining source entity payloads to provide enriched representation of source entities. For example,  {\em artist} and {\em song} entities may need to be joined to produce enriched {\em artist} entities associated with the names of famous {\em songs}. This enrichment operation is done in parallel within each source ingestion pipeline.

\item \emph{Inter-Source Parallelism}: Sources are consumed by knowledge construction via a workflow of parallel pipelines, where each pipeline is internally composed of a number of connected processes, e.g., blocking, pair-generation, and entity matching. The synchronization points across the parallel source pipelines reduce to the fusion operations which need to be conducted on source payloads one at a time. 


\item \emph{Intra-Source Parallelism}: Within each source pipeline, the \emph{Added}, \emph{Updated}, and \emph{Deleted} payloads are processed in parallel. The \emph{Added} payload needs to be fully linked, which requires running all linking pipeline stages. On the other hand, \emph{Updated}/\emph{Deleted} payloads contain entities that are previously linked, and so we only need to lookup their links in the current KG, and perform object resolution operations. The volatile properties payload of a given source are processed by performing a partition overwrite of the KG after the added/deleted payloads  are fused with current KG.
\end{itemize}

 \begin{figure}[t]
        \centering
      \includegraphics[width=1\columnwidth]{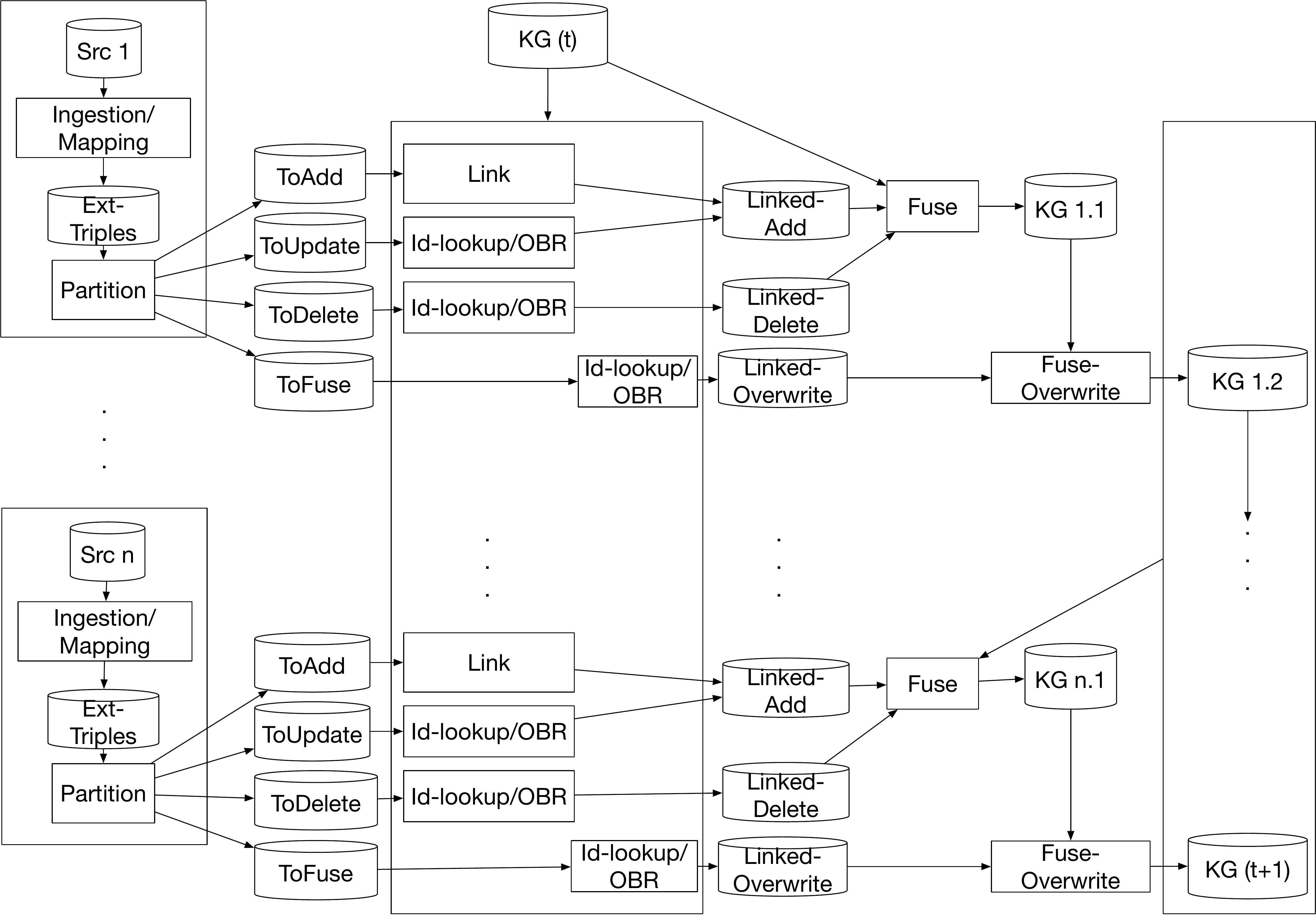}
      \caption{A parallel knowledge graph construction pipeline.}
    \label{fig:par_construction}
\end{figure}

%

\autoref{fig:par_construction} shows the architecture of parallel knowledge graph construction. Source datasets are processed by different pipelines, and synchronization happens during fusion. For each source, the \emph{ToAdd}, \emph{ToUpdate} and \emph{ToDelete} payloads are processed in parallel to incrementally generate the triples to be fed into fusion. When fusion input is ready, the source payloads are fused with the KG and entity links are updated. The \emph{ToFuse} payload of volatile properties is fused with the current KG after the previous source payloads are completely fused. This leverages an optimized fusion path, enabled by maintaining graph partitioning over volatile triples of each source, which  allows overwriting that source partition in the KG with the new updates, without performing expensive joins. 

%% file: sections_sigmod/compute.tex
\section{Knowledge Graph Query Engine}\label{sec:compute}

\begin{figure}[t]
        \centering
      \includegraphics[width=0.95\columnwidth]{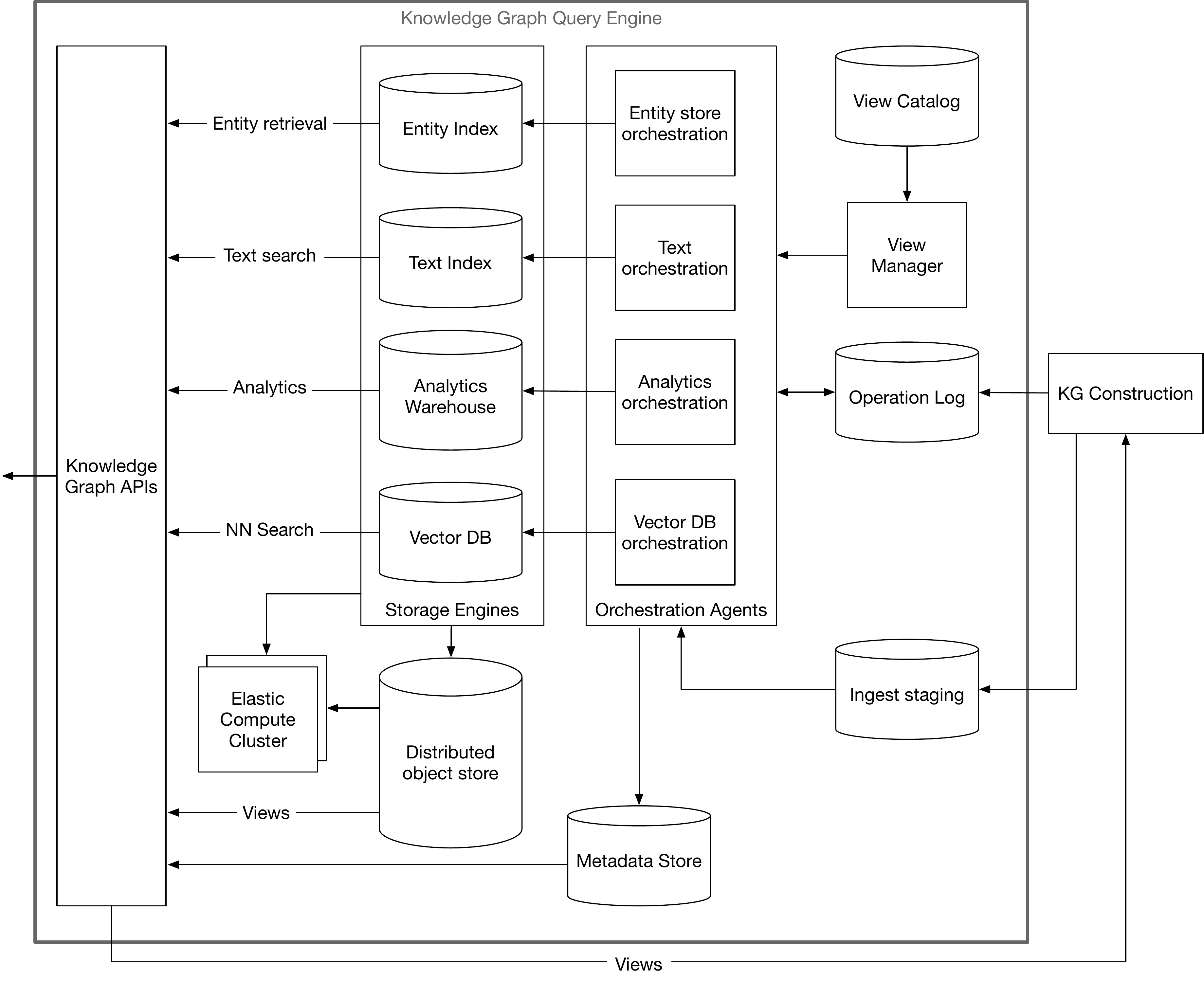}
      \caption{The architecture of the KG Query Engine.}
    \label{fig:kge}
\end{figure}

The Knowledge Graph Query Engine (or \emph{Graph Engine}) serves three purposes within \akp: it is the primary store
for the KG, it computes \emph{knowledge views} over the graph, and it exposes query APIs for graph consumers.
A federated polystore approach~\cite{stonebreaker2015polystore} is used to support the wide variety of workloads against the graph, both in view computation and
query APIs. 
Our workloads include incrementally maintaining KG views, graph learning algorithms,
graph analytics, low-latency entity retrieval, full-text search with ranking, and nearest neighbour search. With such a 
diversity in workloads, specialized engines are required to provide high-quality solutions for each of these verticals. At the same time, we must 
coordinate updates across these engines to ensure consistency of the KG.
An overview of this architecture is shown in Figure~\ref{fig:kge}.

\subsection{Knowledge Graph Storage}
As the primary store for the graph, the Graph Engine is responsible for managing the data lifecycle of the graph as it is updated.
This workload includes updating various indexes across multiple storage engines in a consistent way and maintaining graph versions for analytics.
A distributed shared log is used to coordinate continuous ingest, ensuring that all stores eventually index the same KG updates in the same order.
The log is durable and fault-tolerant, ensuring operations are not lost under a variety of failure scenarios.
An extensible data store orchestration agent framework allows simple integration of new engines, allowing the platform to onboard new workloads
and prototype new storage and compute engines with reasonably small engineering effort. Orchestration agents encapsulate all of the store specific
logic, while the rest of the framework is generic and does not require modification to accommodate a new store type.

The KG Construction pipeline described in Section~\ref{sec:construction} is the sole producer of data.
Data payloads are staged in a high throughput object store and ingest operations are written to a durable operation log
for data ingest. Orchestration agents then process ingest operations in order, ensuring that all stores eventually derive
their domain specific views of the KG over the same underlying base data. Log sequence numbers (LSN) are used as a
distributed synchronization primitive. Orchestration agents track their replay progress in a meta-data store, updating the LSN
of the latest operation which has successfully been replayed on that store. This information allows a consumer to determine the freshness
of a store, ie., that a store is serving at least some minimum version of the KG.
%
%

\subsubsection{Stores and Compute Engines}

The analytics engine is a relational data warehouse that stores the KG extended triples produced by KG construction.
This engine is used for a number of analytics jobs, and generates various subgraph and schematized entity views for upstream tasks (see Section~\ref{sec:views}).
The engine is read optimized, and therefore updates to the engine are batched for performance. 

\subsection{Knowledge Graph Views}
\label{sec:views}
In our experience, most clients want to consume a derived view of the KG rather than the raw graph in its entirety.
Incremental view maintenance is a well studied problem in database literature~\cite{zhuge1995view} .
We adopt a very general definition of a view in our system. A view can be any transformation of the graph,
including sub-graph views, schematized relational views, aggregates, or more complex computations such as iterative algorithms (e.g,. Pagerank)
or alternative graph representations (e.g., vector embeddings).
In all cases, we want to manage the lifecycle of KG views alongside the KG base data itself.
These operations include materializing the views when a new KG is constructed, and incrementally maintaining the views
(when possible) as the KG is updated. Views may specify different freshness SLAs for the Graph Engine to maintain.

View definitions are scripted against the target engines' native APIs. The definitions include procedures for creating and dropping the view,
as well as a procedure for updating the view given a list of changed entity IDs. These definitions are maintained in a central view catalog, 
along with a list of view dependencies. Execution of the view dependency graph is coordinated by the View Manager interacting with the 
Orchestration Agents using a common API over a central message bus. This API is independent of the specific engine which simplifies
extending the platform with new stores.

\begin{figure}[t]
        \centering
      \includegraphics[width=0.8\columnwidth]{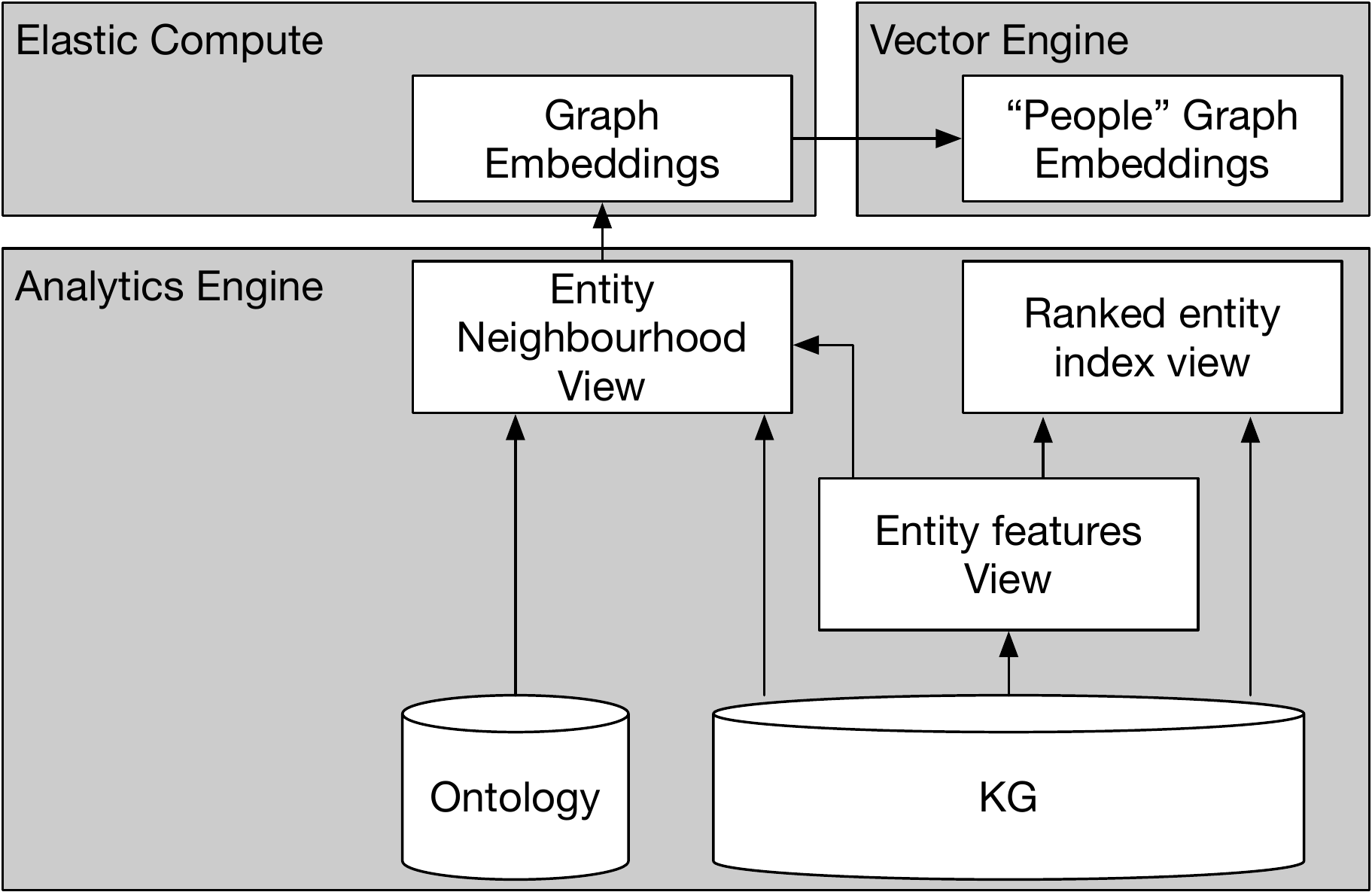}
      \caption{Example of view dependencies}
    \label{fig:view-example}
\end{figure}

As an example (Figure~\ref{fig:view-example}), we use the analytics warehouse to produce a feature view over all entities. These features
are useful for various ranking and machine learning tasks. A ranked entity index view then combines textual references to entities (e.g., names
and aliases) with scoring features to produce an indexible ranked entity view. Independently, an entity neighbourhood view incorporates entity
features in a view that is used to learn graph embeddings. By sharing the construction of the entity features view in the creation of both the 
entity neighborhood and ranked entity index view, we save greatly on overall execution time. Such practices are standard in multi-query
optimization \cite{chakravarthy1986multiple,sellis86global}.
In a production view dependency graph, we found a 26\% run-time improvement when utilizing view dependencies to reuse common views.


Figure~\ref{fig:view-example} also includes an example of cross-engine view dependencies. 
Cross-engine views are orchestrated by the View Manager, including lifecycle of intermediate artifacts. 
In this example, the entity neighborhood view
computed in the analytics engine is consumed by the elastic compute framework where graph embeddings are learned. Those embeddings
are then indexed in a vector database, where an attribute filter on entity type can be used to produce a subset of ``people'' embeddings.

\begin{figure}[t]
        \centering
      \includegraphics[width=0.85\columnwidth]{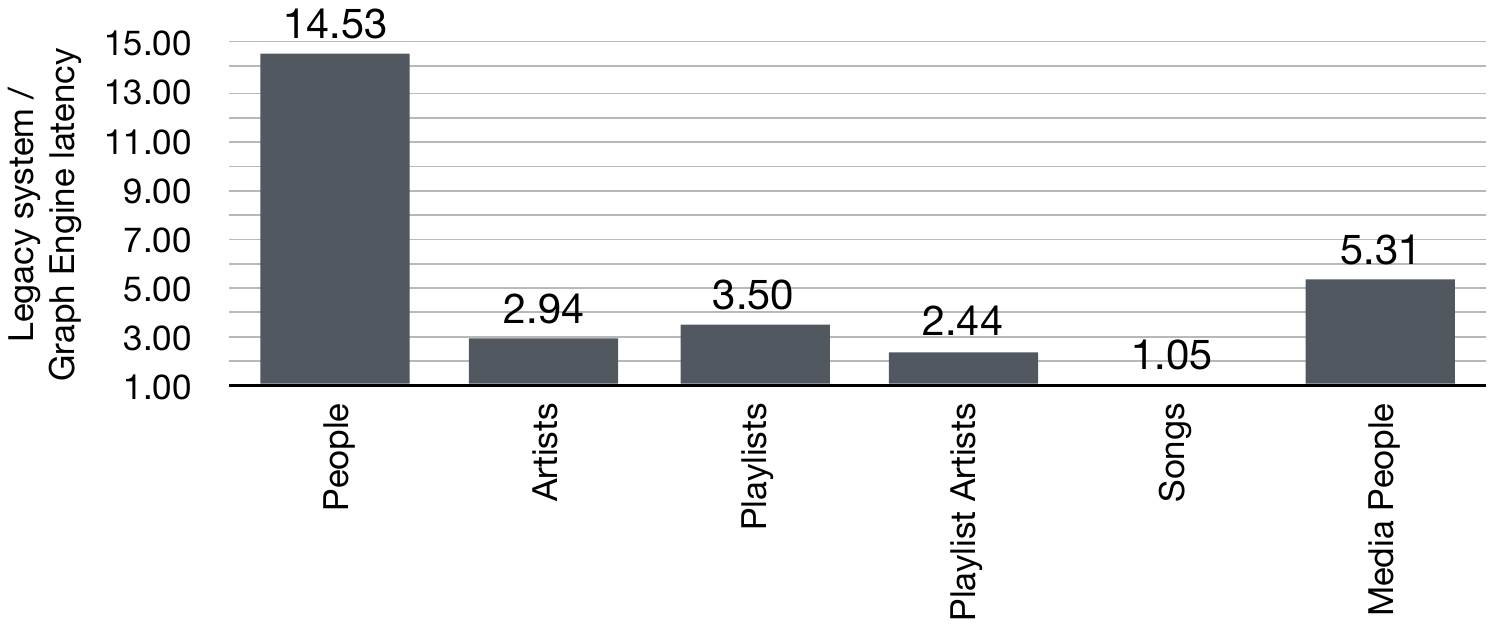}
      \caption{A comparison between \akp's Graph engine view computation vs. a legacy approach.}
    \label{fig:view-perf}
\end{figure}

Having a variety of specialized storage engines not only permits a variety of view definitions (from relational to learned embeddings), but
also allows optimized view implementations using the best engine for each task.
Figure~\ref{fig:view-perf} shows the performance results of using the Graph Engine's Analytics Store to compute a set of views used in
a production scenario. The graph illustrates relative performance gain compared to a legacy implementation of the views as custom Spark jobs.
These views compute entity-centric schematized relational views for a variety of entity types shown on the x-axis. The optimized join
processing in the Analytics Store yields an average of $5x$ performance improvement with up to $14x$ in the best case for these join-heavy
view definitions. The lowest increase was the ``Songs'' view which had only a $5\%$ increase. No views had a performance decrease.  
In these experiments, the legacy system uses nearly ten times the amount of hardware. It is worth noting that Spark-based execution is well
suited for other types of views (e.g., highly parallelizable tasks, machine learning tasks or views with a 
large amounts of string manipulation).
These results highlight the importance of the polystore approach, allowing the best compute engine to be used for each view.

\subsection{Entity Importance}\label{sec:ei}
Many KG use cases involve ranking entities. In some situations, external signals of popularity provide provide a
effective ranking signal. For example, song plays, search frequency, or UI engagement on entities. However these types of
popularity metrics tend to cover head entities and are weaker or absent for less popular entities. Applications of the KG that need to rank all entities require a metric that covers tail and torso entities as well as head entities.

There are a number of structural signals in the graph that can be used to estimate the \emph{importance} of an entity, based
on its connectivity in the graph. Simple metrics like in-degree and out-degree can contribute to an importance score. The intuition
here being that the more we know about an entity, the more important it must be. However, entities from certain sources may
have many more properties than entities from other sources, so degree alone is not sufficient as it would bias entities occurring
in particular sources. We incorporate four structural metrics to score the importance of an entity in the graph.
In-degree, out-degree, number of identities, and Pagerank~\cite{brin1994anatomy} in the graph. Number of identities corresponds to the number of
sources that contribute facts for the entity. Pagerank is computed over the graph, recursively scoring the importance of an entity node
based on its connectivity, and the connectivity of its neighbours. We then aggregate these metrics into a single score representing
the importance of the entity based on graph structure.


The computation of entity importance is modelled as a view over the KG, computed by the analytics engine. The
view is registered with the view automation described in Section~\ref{sec:views} and is automatically maintained as the graph changes.

%% file: sections_sigmod/livegraph.tex

\section{The Live Graph}
\label{sec:livegraph}
 \begin{figure}[t]
        \centering
      \includegraphics[width=0.8\columnwidth]{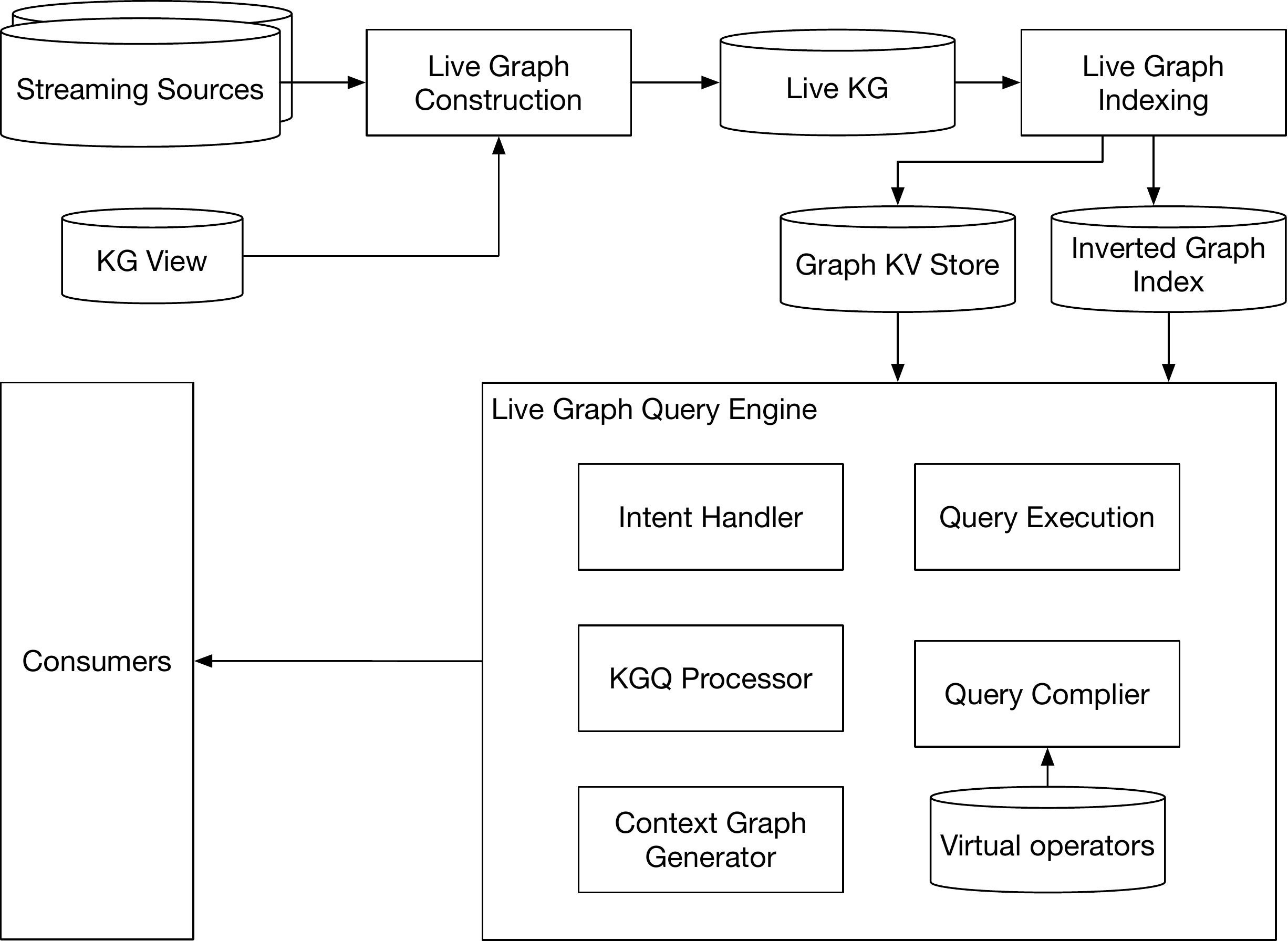}
      \caption{The Live Knowledge Graph.}
    \label{fig:live-graph}
\end{figure}
Our KG is built from a variety of sources that contribute \emph{stable} knowledge. We compliment this data with
\emph{live} sources of knowledge that contribute temporal facts in real-time. Such sources include sports scores, stock prices,
and flight information. 

The \emph{live KG} is the union of a view of the stable graph with
real-time live sources. The live graph query engine is highly optimized for low-latency graph search queries, and is geo-replicated for 
serving locality. This engine powers real-time search over the KG for various use cases like open-domain 
Question Answering, KG cards, sports scores and other domain specific experiences.
The live KG engine handles billions of queries daily while maintaining strict latency SLAs for interactive systems.
An overview of this architecture is shown in Figure~\ref{fig:live-graph}.

\subsection{Live Graph Construction}

Live KG Construction is the process of building and linking a KG that integrates a view of 
stable knowledge with live streaming sources, such as sports scores, stock prices, and flight data.
Live sources do not require the complex linking and fusion process of our full KG construction pipeline, ie., sports games,
stock prices, and flights are uniquely identifiable across sources and do not have the inherent ambiguity that requires linking 
different mentions of the same sports game, stock reference, or flight. These sources do contain potentially ambiguous references 
to stable entities which we want to link to the stable graph. For example, we want to resolve the references in a sports game to the 
participating teams, the stadium or venue, and the city where the game takes place. We utilize the Entity Resolution service described
in Section~\ref{sec:nerd} to resolve text mentions of entities to their stable entity identifiers.

The result of Live Graph Construction is a KG that includes continuously updating streaming data sources who's entity 
references are linked to the stable graph. This design allows us to build applications that query streaming data (e.g., a sports score) while using
stable knowledge to reason about entity references. 

The live KG is indexed using a scalable inverted index and key value store. Both indexes are optimized for low latency retrieval under high degrees
of concurrent requests. The indexes are sharded and can be replicated to support scale-out. This design allows tight control over the load an individual
index server supports. 

\subsection{Query Execution}
The Live KG Query Engine process ad-hoc structured graph queries and \emph{query intents} which consist of a target
intent and arguments. The engine also maintains query context to support mutli-turn interactions. The architecture we follow is similar to standard dialogue systems from both academia and industry~\cite{alexa_dialogue,john2017ava,fast2018iris}.

\paragraph{Live Graph Queries}
The Live KG Query Engine supports ad-hoc structured graph queries against the KG
with strict latency SLAs in order to support interactive use cases like Question Answering. Clients can specify queries 
using a specially designed graph query language called \emph{KGQ}. KGQ is expressive enough to capture the semantics of natural language (NL) queries coming from 
our front end search interfaces, while limiting expressiveness (compared to more general graph query languages) in order to bound query performance. The
queries primarily express graph traversal constraints for entity search, including multi-hop traversals.
%
KGQ is an extensible language, allowing users to implement \emph{virtual operators}. Virtual operators allow complex expressions to be encapsulated as
new operators, facilitating easy reuse of complex expressions across different use cases.

The Live Graph Query Engine compiles queries into a physical execution plan. The engine allows pluggable storage back-ends and makes use of both inverted
indexes and key-value stores for live KG query evaluation. A number of execution optimizations are used, including operator push-down and 
intra-query parallelism. Combining this execution with the scalability and performance of the underlying inverted index and key value store, as well as caching, 
allow the engine to achieve 95$^{th}$ percentile query latencies of less than 20s of milliseconds on production workloads.

\paragraph{Query Intent Handling}
In addition to KGQ execution, the Live Graph Query Engine also supports a comprehensive query intent handler. 
The intent handler processes annotated natural language queries by routing intents to potential KGQ queries based on the annotations.
For example, the queries
``\textit{Who is the leader of Canada?}'' and 
``\textit{Who is the leader of Chicago?}''
share the high-level query intent, each with their respective arguments.
``\textit{HeadOfState(Canada)}'' and
``\textit{HeadOfState(Chicago)}''
Despite having the same intents, the graph queries needed to answer these two queries are different. In the first case, we want to find the entity that
is the \emph{prime minister} property of the entity argument Canada. In the second, we want the \emph{mayor} property of the entity Chicago. Intent routing solves this problem
by choosing the correct execution based on the semantics of the entities, i.e., there is no mayor of Canada or prime minister of Chicago, only one interpretation
is meaningful according to the semantics encoded in the KG.

\paragraph{Query Context}
The Live KG Query Engine also maintains a context graph and intents from previous queries to support follow-up queries. Query sequences, such as
{\scriptsize\begin{align}
Query&:& \textit{Who is Beyonc\'e married to?}\\
Intent&:& \textit{SpouseOf(Beyonc\'e)}\\
Answer&:& \textit{ $\rightarrow$ Jay-Z }\\ 
Query&:&\textit{How about Tom Hanks?}\\
Intent&:&\textit{SpouseOf(Tom Hanks)}\\
Answer&:&\textit{ $\rightarrow$ Rita Wilson }\\ 
Query&:&\textit{Where is she from?}\\
Intent&:&\textit{Birthplace(Rita Wilson)}\\
Answer&:&\textit{ $\rightarrow$ Hollywood }
\end{align}}
maintain context graphs in order to reference previous query intents, or entities from previous queries. This context can be referenced by the engine to 
bind parameters for query execution. We see in the example, that the intent on line 5 is taken from the previous interaction. On line 8, the intent is derived
from the query issued on line 7, but the argument for the intent is pulled from the context graph from the previous interaction.

\subsection{Live Graph Curation}

Our end-user experiences depend on the knowledge platform producing a high quality KG. Quality not only refers to the accuracy of linking and fusing
knowledge, but also to the quality of the data itself.
The quality of source data can vary widely depending on the source. Some sources may occasionally contain errors, and some sources are subject to
vandalism from community edits. To address this, we integrate a \emph{human-in-the-loop} curation pipeline. Facts containing potential errors or vandalism
are detected and are quarantined for human curation. A team can block or edit particular facts or entities using custom built curation tooling. These curations
are treated as a streaming data source by the live graph construction which allows us to \emph{hot fix} the live indexes directly when the curation process
identifies an error. The curations are also sent to the stable KG construction as a source, so that corrections are incorporated into the stable graph.

%% file: sections_sigmod/services.tex

\section{Graph Machine Learning}\label{sec:ml}

\subsection{Neural String Similarities}\label{sec:nss}
Accurate duplicate detection is a key requirement during KG construction. We provide a library of similarity functions for different data types that developers can use to obtain features when developing matching models. Beyond deterministic similarity functions (e.g., Hamming distance, Jaccard similarity, and Edit Distances~\cite{integration_book}), \akp offers several learned string similarity functions that help boost the recall of matching models by capturing semantic similarities such as synonyms~\cite{bilenko2003adaptive, cohen2003comparison}. These learned similarity functions can be used out-of-the-box to featurize the input to matching models that are used during KG construction.


\akp's learned similarity functions rely on neural network-based encoders that map a sequence of characters into high-dimensional vectors~\cite{kim2016character}. Given the vector representations of two strings we compute their similarities by taking the cosine similarity of their vector representations. If trained with appropriate data these neural encoders can yield string similarity functions that are capable to go beyond typos and can capture synonyms (e.g., they can capture that ``Robert'' and ``Bob'' are similar). To ensure homogeneity of these representations and capture the structural difference across names of different entity types, we learn different neural string encoders for different types of strings, e.g., human names, location names, music album titles etc. 

For training we use \emph{distant supervision}~\cite{mintz2009distant}. We bootstrap the information in the KG to obtain a collection of training points for each of the string encoders. Entities in the KG are associated with multiple aliases and names. We use this data to obtain examples of pairs of strings that should be similar. Simple augmentation rules based on typos are also be used to generate positive examples. Such \emph{data augmentation} practices are standard in training deep learning models~\cite{wei2019eda, heidari2019holodetect}. To generate negative examples, we leverage the entities in the graph to generate negative examples (i.e., pairs of string that should not be similar) by using the names and aliases of entities that are not linked. These examples are used to form a \emph{triplet} loss that is then used to train the encoder for each string type. The learned encoders and corresponding similarity functions are transferable and are currently deployed in use cases beyond KG construction. In cases where typos and synonyms are present, we have found that using these learned similarity functions can lead to recall improvements of more than 20 basis points.

\subsection{Entity Recognition and Disambiguation}\label{sec:nerd}
Named entity recognition and disambiguation (NERD) is the problem of identifying text mentions of named entities in unstructured or semi-structured data and disambiguating them against entities in a KG or standardized vocabulary~\cite{nadeau2007survey, orr2020bootleg, nguyen2014aida, burdick2016declarative, lample2016neural, nadeau2007survey, mulang, Yamada2019GlobalED}. For example, given the sentence `We visited Hanover and Dartmouth' or the record \texttt{<Dartmouth, located\_in: Hanover>} we want to resolve the mention ``Hanover'' to \texttt{Hanover, New Hampshire} and not to the more popular \texttt{Hanover, Germany}. 

\begin{figure}[t]
        \centering
      \includegraphics[width=\columnwidth]{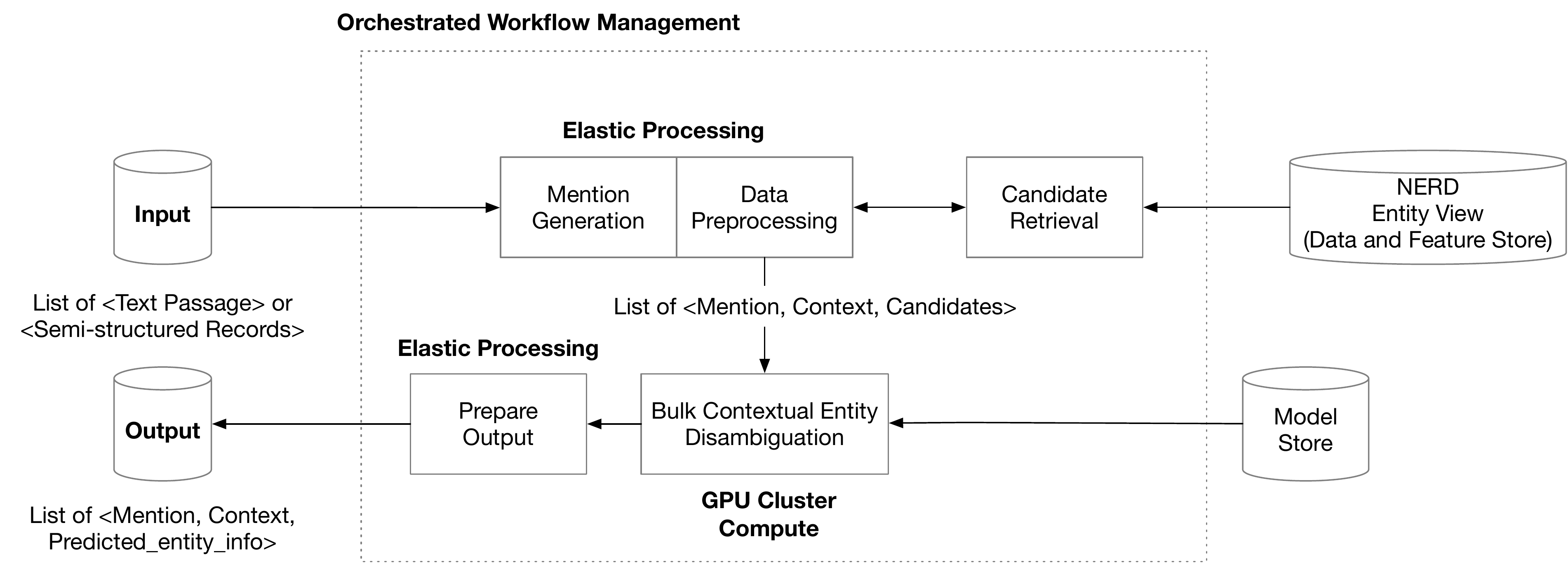}
      \caption{An overview of the batch deployment architecture and main components of \akp's NERD stack.}
    \label{fig:nerd_arch}
\end{figure}

\akp provides a complete NERD stack, which is used to implement the \emph{object resolution} during KG construction (see Section~\ref{sec:construction}) but also powers a number of additional use cases where annotating or enriching text-based data with information from the KG is required. We use an elastic deployment for large batch jobs and a high performant low-latency variant for online workloads. Figure~\ref{fig:nerd_arch} shows a high-level diagram of the batch deployment and the main components of the NERD stack. 

We treat entity disambiguation as an entity linking problem~\cite{burdick2016declarative}. A key requirement in \akp is our ability to correctly disambiguate \emph{tail} (i.e., less popular) entities. In this case, one cannot rely only on string similarities between the mention and entity names in the graph but needs to reason about the context (e.g., surrounding text or other fields in a structured record) that a mention appears in. Such context can carry information about the relationships or the semantic type of the entity that the mention refers to and can be compared against information in the KG to improve the accuracy of named entity disambiguation~\cite{orr2020bootleg, Yamada2019GlobalED, mulang}. To this end, we create a view using the Graph Engine described in Section~\ref{sec:compute} that summarizes our knowledge for each entity in the KG, i.e., its aliases, entity types, relationships, types of its neighboring entities, and reason about similarities between the context of a mention and these entity summaries. We refer to this view of entity summaries as \emph{NERD Entity View}. Given a mention and the relevant context, our goal is to find if there exists any record in the NERD Entity View that is a ``match'' of the mention in the input. The first step is to identify candidate entities that are likely to be matches to the mention. Then, we compute a matching score for each of the returned candidates and identify if there is a record in the NERD Entity View that matches the input mention with high-confidence.

\paragraph{NERD Entity View} The goal of each record in the NERD entity view is to provide a comprehensive summary that can act as a discriminative definition for each entity in the KG. Each entry in the NERD Entity View is a record with attributes that contain information about: 1) the name and aliases of the entity in different locales, 2) the different types from the KG ontology that are associated with the entity (e.g., `human', `music artist', `academic scholar' etc), 3) a text-based description of the entity if available, 4) a list of important one-hop relationships that the entity participates in, 5) the entity types of important neighbors of the entity, and 6) the entity importance scores computed by the Graph Engine (Section~\ref{sec:ei}). This comprehensive summary of each entity in the KG provides opportunities to identify cases where information in the NERD Entity View overlaps with information in the context and hence perform more accurate disambiguation. For example, given that the NERD Entity View for \texttt{Hanover, New Hampshire} includes the relationship \texttt{<Dartmouth College, located\_in, Hanover>}, we can accurately identify that the mention ``Hanover'' in the context of the sentence `We visited downtown Hanover after spending time at Dartmouth' refers to \texttt{Hanover, New Hampshire} and not \texttt{Hanover, Germany}. The NERD Entity View is computed using the Graph Engine, which guarantees the its freshness via incremental updates as new facts and entities are ingested in the KG.

\paragraph{Candidate Retrieval} Candidate retrieval can be viewed as a parallel to blocking in entity linking. In this step we rely on the similarity between the input entity mention and the name and alias fields of the records in the NERD Entity View to find likely matches. To go beyond exact matches, we use the neural string similarity functions described above. We also allow information on admissible entity types to be used to further improve precision---we make use of Entity Type information during Object Resolution in KG Construction where the attribute-value to be disambiguated is accompanied by an entity type (see Section~\ref{sec:usecases}). In the presence of constraints on computational resources or tight latency requirements, we rely on entity importance to prioritize candidate comparison and limit the scope of entity disambiguation to popular entities. Overall, given a limit of $k$-candidates the goal of candidate retrieval is to optimize recall by \emph{pruning} the domain of possible matches given the extreme and ever-increasing number of entities in the KG. This approach is inspired by our prior work on HoloClean~\cite{holoclean, aimnet} where pruning was shown to be critical for accurate data cleaning and imputation over extremely large domains. 

\begin{figure}[t]
        \centering
      \includegraphics[width=\columnwidth]{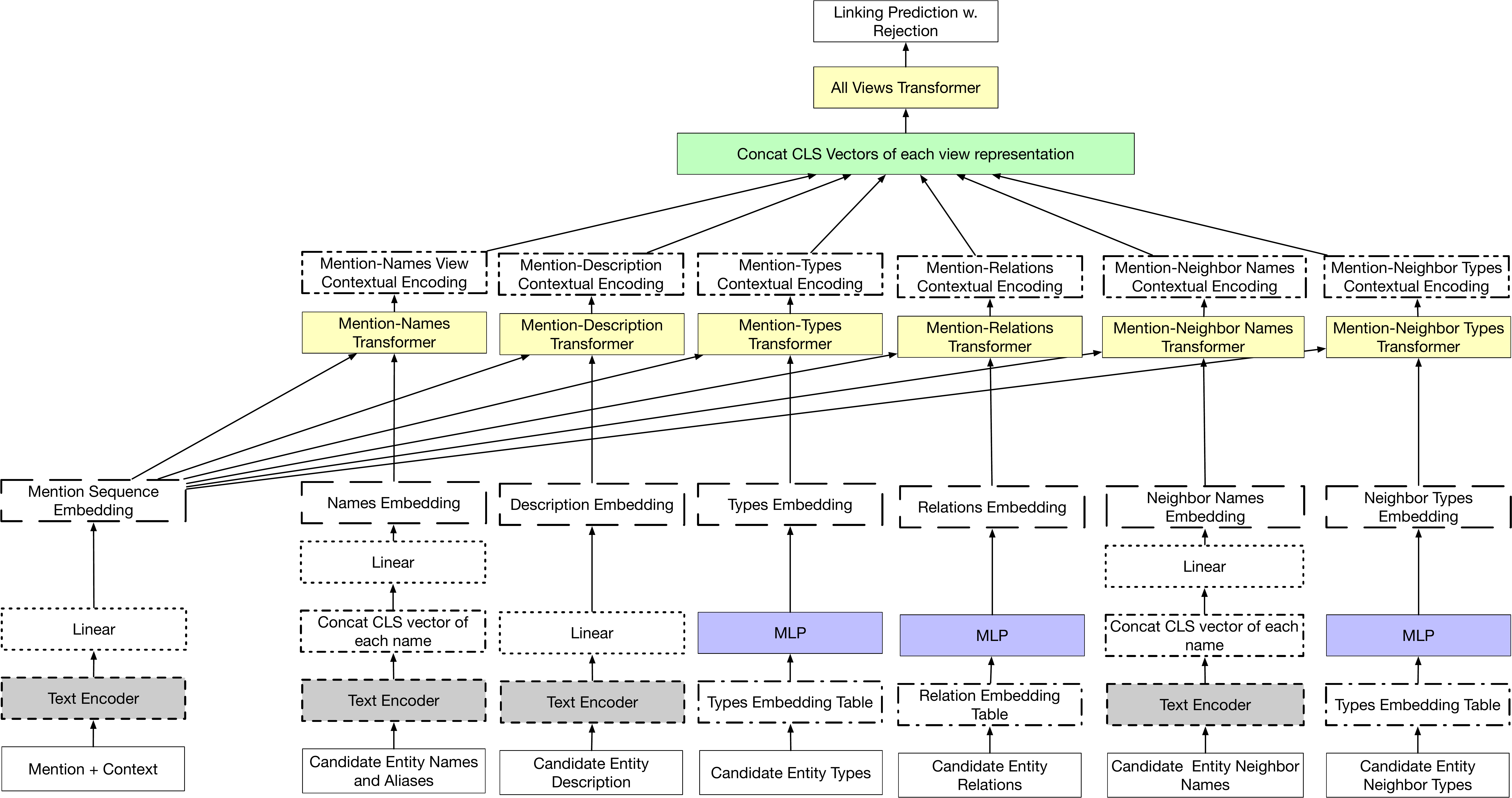}
      \caption{The Transformer-based Contextual Entity Disambiguation model in the NERD stack.}
    \label{fig:nerd_reranker}
\end{figure}

\paragraph{Contextual Entity Disambiguation} The last step of the NERD stack is responsible for determining which of the entity candidates (if any) is the most probable to be referenced in the input mention. We cast Entity Disambiguation as a \emph{classification problem} over the space of available candidates with an additional rejection mechanism, i.e., we allow rejecting all input candidates as not good options. To enable classification over sets of candidates with variable input size and provide the opportunity for rejection we rely on a \emph{one versus all} version of multi-class classification~\cite{rejection}. We also follow a neural network architecture that is similar to state-of-the-art named entity disambiguation models~\cite{orr2020bootleg, Yamada2019GlobalED} and models that jointly encode graphs and text~\cite{chen-etal-2020-kgpt,wikigraphs}. Specifically, the model we use to perform this classification task is a contextual, transformer-based deep neural network that leverages the Attention mechanism~\cite{attention} to reason about the similarity between the input context and the different attributes in the NERD Entity View records. A diagram of our model and approach for Entity Disambiguation is shown in Figure~\ref{fig:nerd_reranker}. All models used in the NERD stack are trained offline via weak-supervision procedures that combine a collection of text data annotated with entity tags, manually curated query logs, and text snippets generated by applying templates over a selection of facts present in the KG. While these models are re-trained at regular intervals to ensure no accuracy degradation, entity additions are reflected by updating the NERD Entity View.

\subsection{Knowledge Graph Embeddings}\label{sec:kge} \akp uses modern ML over graph-structured data to enable functionalities such as \emph{fact ranking}, \emph{fact verification}, and \emph{missing fact imputation}. Fact ranking seeks to provide an importance-based rank over instances of high-cardinality entity predicates. For example, given a list of multiple occupations such as `singer', `television actor', `songwriter' for an entity, we want to determine the \emph{dominant} occupation to enable more engaging experiences for our users. Fact verification seeks to identify facts in the graph that might be erroneous, i.e., correspond to outliers, and should be prioritized for auditing. Finally, missing fact imputation can expand the KG with facts that are inferred via transitivity or other structure-based inferences. Beyond rule-based solutions, we also rely on ML link-prediction approaches that leverage \emph{knowledge graph embeddings} to provide a unified solution to these problems.

KG embeddings use machine learning models to assign each entity and predicate in a KG to a specific continuous vector representation such that the structural properties of the graph (e.g., the existence of a fact between two entities or their proximity due to a short path) can be approximated using these vectors. Given a subject entity $s$ and a predicate $p$ in the KG, one can use a learned model that takes as input the embeddings $\theta_s$ and $\theta_p$ of entity and the predicate to obtain a vector $f(\theta_s, \theta_p)$ that can be used to find possible objects for this fact via \emph{vector-based similarity search} between $f(\theta_s, \theta_p)$ and the embeddings of all entities in the KG. \akp leverages this similarity search to unify the tasks of fact ranking, fact verification, and missing fact imputation. In the presence of a known object entity $o$ that forms the fact <$s$, $p$, $o$>, we use the similarity between $f(\theta_s, \theta_p)$ and the embedding $\theta_o$ to obtain an importance score for that fact and leverage that score during both fact ranking and fact verification. On the other hand, in the absence of an object for the tuple <$s$, $p$> we perform nearest neighbor search by leveraging the Vector DB component of the Graph Engine to identify potential candidate objects that complete the fact. 


Since different embedding models capture different structural properties of KGs, we do not rely on a single model but we opt for a generalizable architecture that allows us to train multiple embedding models including standard models like TransE~\cite{transe} and DistMult~\cite{distmult}. To prepare the necessary data for training, we leverage the relational store of the Graph Engine and register a specialized view that filters unnecessary metadata facts from the KG to retain only facts that describe relationships between entities. We assign training of each embedding model on a separate single-node with multiple-GPUs in our GPU cluster. Finally, the learned embeddings are stored in the Vector DB store of the Graph Engine which provide similarity search functionalities. Given our need to train multiple embedding models over billions of facts and entities, we opt for a single-box multi-GPU training per embedding model to allow for optimized utilization of our GPU resources and leverage the Marius system for training each model~\cite{marius}. 

Training graph embedding models over billion-scale KGs is an extremely memory intensive operation. To learn accurate representations, we need to use high-dimensional vector representations (e.g., 400-dimensional real vectors) for each entity in our graph. Such a representation requires 1600 bytes of storage per node and requires 80 GB (the largest GPU memory) for a small 50 million node graph. Thus, it is necessary to store the learnable parameters in off-GPU memory. Moreover, the memory required to store the learnable parameters for the embedding models exceeds the capacity of available main memory. As such, scaling to graphs of this size requires using either distributed training across multiple GPU-nodes or external memory training. In \akp, we opt for external memory training with the Marius system due to ease of deployment over our GPU cluster. Utilizing the disk memory during training allows us to easily deploy a different instance per multi-GPU node and hence train multiple embedding models without deploying complex scheduling solutions. Training embedding models over the KG with Marius takes one day. On the other hand, we find that competing solutions for scalable graph learning such as DGL-KE~\cite{dgl_ke} and Pytorch BigGraph~\cite{biggraph} either require allocating all GPU resources over the cluster to the training of a single model or present low-utilization of the GPU which leads to the training of these models to span multiple days.

%% file: sections_sigmod/usecases.tex

\section{Use Cases}\label{sec:usecases}

We discuss \akp use cases and the corresponding deployments.

\subsection{Open-Domain Question Answering}\label{sec:siri_qa}


Open-domain question answering seeks answer user questions such as ``What is the tallest mountain in the world?'', or ``Who is the mayor of New York City?'', or even time-sensitive queries such as ``Who's winning the Warriors game?'' The ability of open-domain question answering solutions to answer these questions is dependent on accurate, up-to-date information served from the KG. We describe how question answering leverages \akp to ensure high quality answers are provided in tight SLAs to users.
 \begin{figure}[t]
        \centering
      \includegraphics[width=0.7\columnwidth]{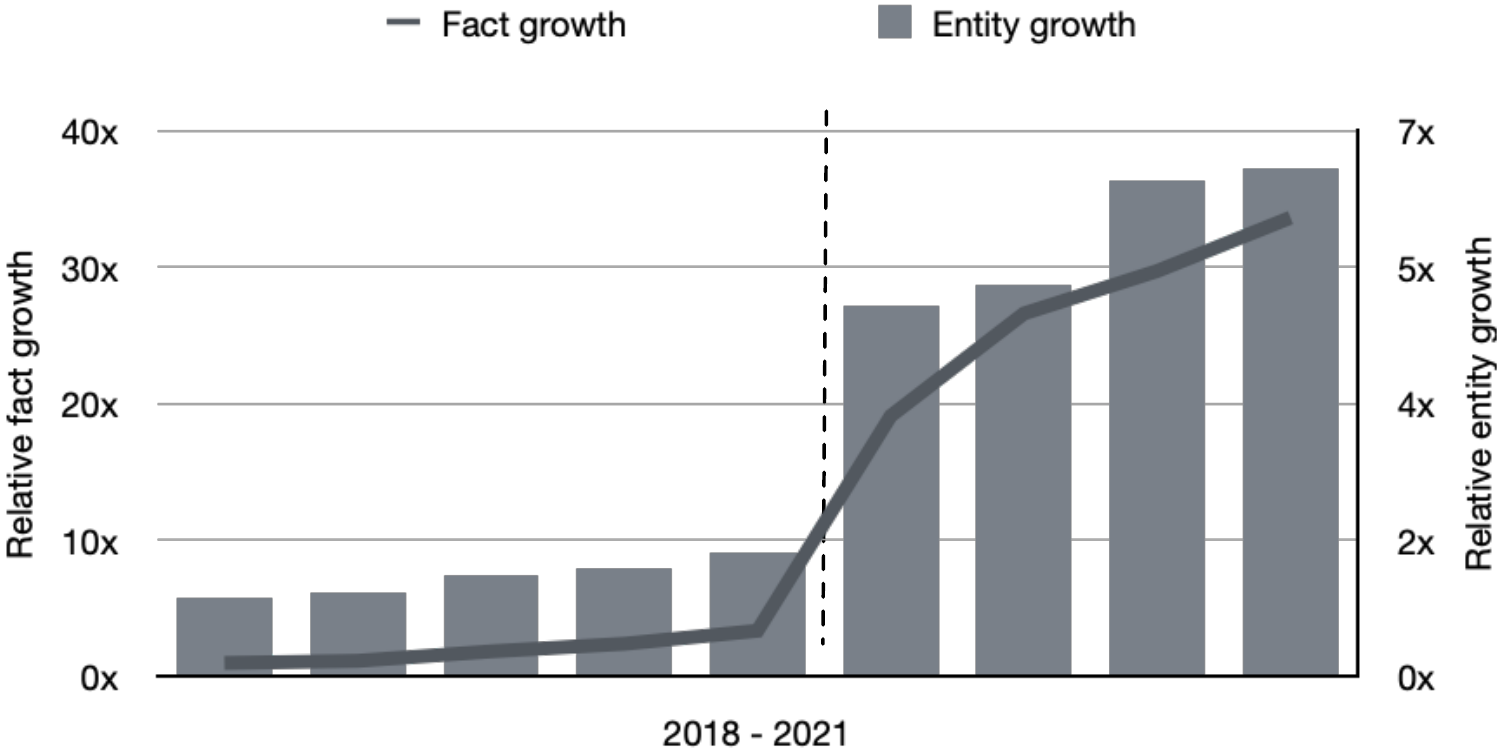}
      \caption{Relative growth of the KG using \akp.}
    \label{fig:kg-growth}
\end{figure}

Natural language understanding and query evaluation are key steps for answering user questions. Critical \akp services contribute to understanding and providing the correct answer including NERD and the Live KG Index. Given a text-based mention of an entity in a user utterance (e.g. ``Joe Biden'' ), we leverage NERD to produce the most likely KG entity (e.g. \texttt{AKG:123}). In parallel, we infer the intent of the user utterance to produce a structured query over the KG (e.g. ``How old is Joe Biden'' yields the query \texttt{ageOf(``AKG:123'')}). The machine-executable query runs over the Live KG Engine to retrieve the correct fact based on the intent and query arguments. In this particular example, we would return the value for the age property for entity \texttt{Joe Biden}.
The Live KG Query Engine powering these queries serves billions of queries per day while maintaining $20ms$ latencies in the $95^{th}$ percentile.

A key challenge in supporting open-domain question answering is ensuring accurate, up-to-date facts in our KG, while expanding the breadth of data available to the query answering stack. Through a combination of multi-source corroboration, fast delta data updates, and targeted fact curation, we support many types of question and answer pairs. The open-domain nature of question answering imposes unique requirements on fact provenance and freshness in our KG. 
The \akp architecture described above allows for the flexibility to support all of these varying workloads to produce a constantly up-to-date and growing KG. Figure~\ref{fig:kg-growth} illustrates the relative growth of facts and entities in the \akg since 2018. The dashed line indicates the point 
at which \akp was introduced. We see over 33$\times$ increase in the number of facts and a 6.5$\times$ increase in the number of
unique entities since the initial measurement. 

\subsection{Entity Cards}\label{sec:entity_cards}

Entity Cards display rich entity-centric information. \akp powers the creation of such cards to provide a diverse set of facts about entities across various domains. Despite Entity Cards being used across different verticals, the common use case of \akp highlights the value of centralizing knowledge construction to provide a consistent, unified experience to users.
For example, when searching for an entity (e.g. ``the singer Billie Eilish''), the KG provides the necessary facts to compile a rich view of the entity including facts about date of birth, age, place of birth, record label, and full name. Relevant entity neighbors around the main entity \texttt{Billie Eilish} are also provided, including her music albums ranked by popularity, social media links, recent news, videos, images and relevant links. Although much of this data is scattered among different sources, \akp ingests and links these data sources to produce a single canonical \texttt{Billie Eilish} entity with all relevant facts. 


Entity cards are also available to vertical applications where entities can for instance be limited to map locations or points of interest. Different vertical use cases leverage specialized KG views to build the appropriate Entity Cards. Such specialized views may require a completely different set of entities and facts to be available in the KG. The scalable, domain agnostic architecture of \akp enables the same pipelines to process both open-domain and domain-specific data to create similar canonicalized views of entities.


\subsection{Semantic Annotations with NERD}\label{sec:annotations}

 \begin{figure}[t]
        \centering
      \includegraphics[width=\columnwidth]{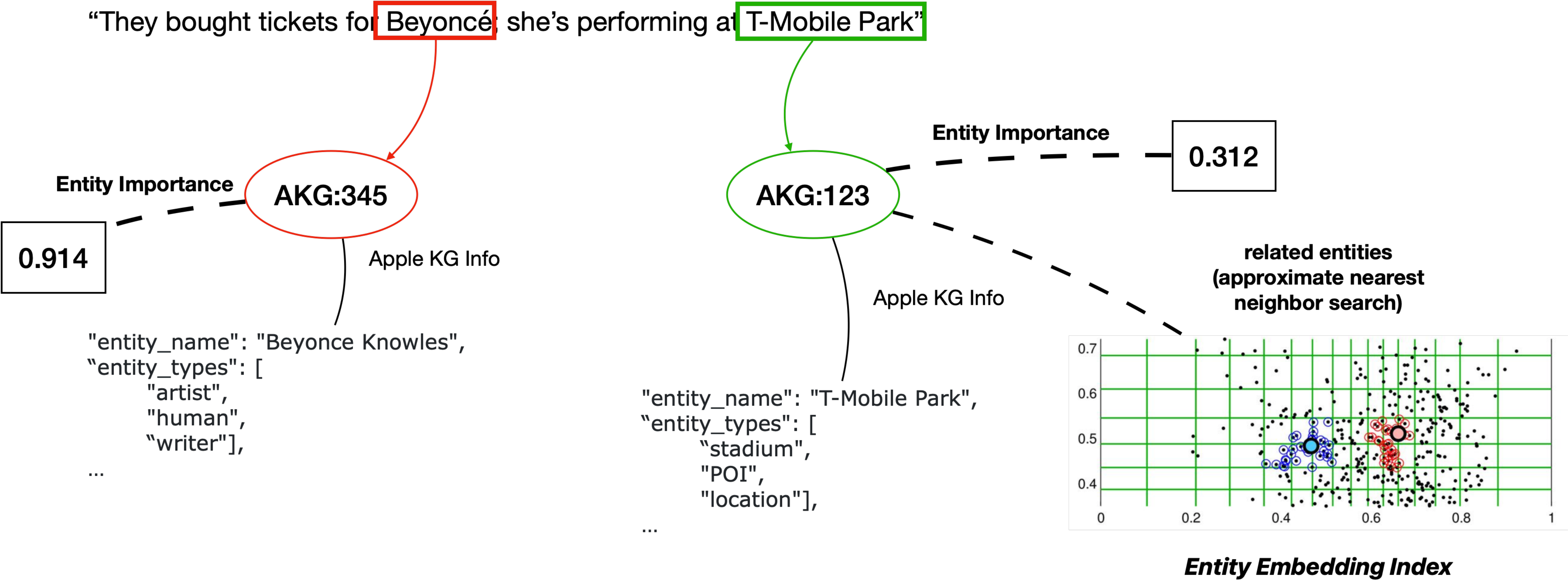}
      \caption{Semantic annotations from the KG with NERD.}
    \label{fig:semantic_annotations}
\end{figure}

\akp's NERD is used to power KG construction and to annotate text data with semantic information from the KG. An example of such annotations is shown in Figure~\ref{fig:semantic_annotations} where short text highlights are augmented with information from the KG using NERD. Once NERD has disambiguated text mentions to entities, \akp can provide additional information such as entity importance scores, embedding-based representations, and related entities from the KG. This semantic metadata enables content understanding and provides a useful signal for content categorization and search.

\begin{figure}[t]
        \centering
      \includegraphics[width=\columnwidth]{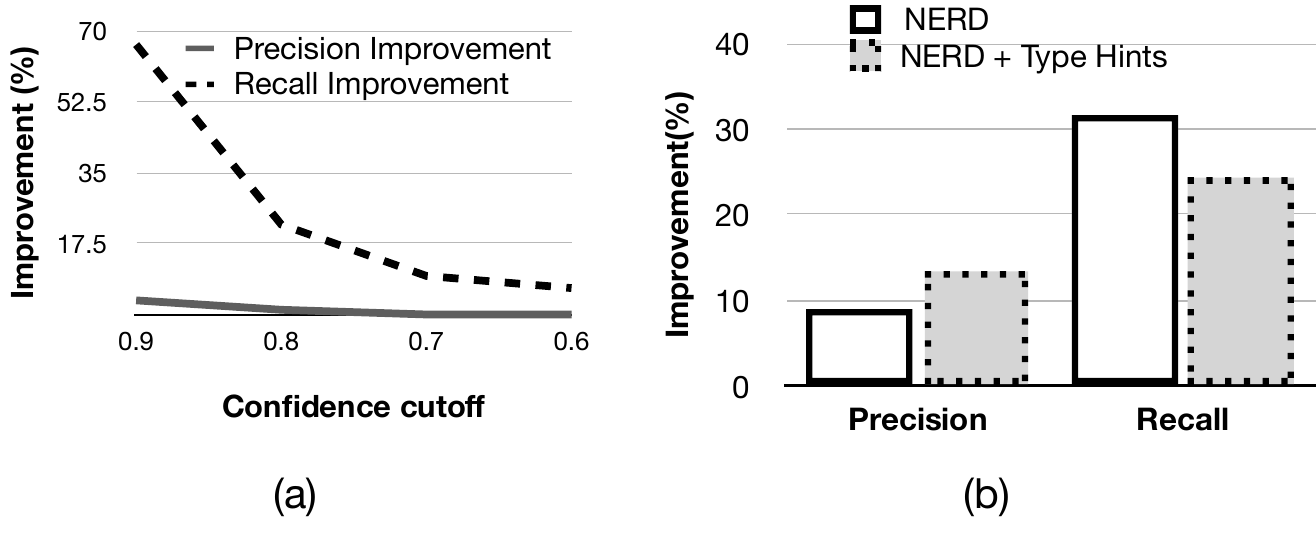}
      \caption{NERD vs an existing deployed method for (a) text annotations and (b) object resolution.\vspace{-4mm}}
    \label{fig:nerd_eval}    
\end{figure}

NERD's use cases are span two groups: 1) annotation of text documents and 2) object resolution. 
For text documents, NERD yields recall improvements while it maintains the same level of precision against an alternative, deployed Entity Disambiguation solution. The main differences between NERD and this approach is that the latter does not leverage the relational information for the entities in the KG but it relies on training data to learn entity correlations and dependencies and encodes this correlations in a neural network. This design promotes high-quality predictions for head entities but not tail entities. Figure~\ref{fig:nerd_eval}(a) shows the relative improvement in precision and recall for different confidence thresholds for accepting or rejecting a prediction. For a confidence level of 0.9 the NERD stack provides a recall improvement of close to 70\%. For lower thresholds the improvements naturally diminish. For high-confidence thresholds i.e., greater or equal than 0.8, NERD also provides precision improvements up to 3.4\%. 

We also find that NERD provides both precision and recall improvements when compared against the aforementioned alternative solution for object resolution in graph construction. We fix the confidence threshold to 0.9 as accurate entity disambiguation is a requirement during knowledge construction. The results are shown in Figure~\ref{fig:nerd_eval}(b). We compare two versions of the NERD stack against the competing solution: Original NERD and a variation of NERD that makes explicit use of entity type hints to obtain higher precision. Recall that entity types of the entity mentions to be disambiguated during object resolution correspond to known types in our ontology. As shown, NERD with type hints yields a precision improvement of around 10\%. It also yields a recall improvement of around 25\% against the alternative solution.

%
%
%

%% file: sections_sigmod/related.tex

\section{Related Work}\label{sec:related}


Knowledge graphs became prevalent with seminal projects such as DBPedia~\cite{dbpedia}, Freebase~\cite{freebase}, KnowItAll~\cite{knowitall}, WebOfConcepts~\cite{webofconcepts}, and YAGO~\cite{yago}. These efforts were followed by community-driven projects such as WikiData~\cite{wikidata} and projects that explored the application of modern ML to scale the construction of large-scale KGs by extracting information from unstructured data~\cite{deepdive_sigmod_record, nell, knowledgevault}. KGs have also become a key asset in industrial applications, including search, analytics, and recommendations. Industrial KGs span both general purpose and vertical deployments and more~\cite{industry_kgs, satori, yang2019aligraph, dong2018challenges}.

KG construction spans multiple technical areas in Data Management and Artificial Intelligence. Techniques developed for data integration~\cite{integration_book, lenzerini2002data}, data cleaning~\cite{ilyas2019data}, view maintenance~\cite{zhuge1995view} and large-scale graph data processing and analytics~\cite{aggarwal2010graph, jindal2014vertexica, fan2015case} are critical to ensure the accurate and scalable construction of KGs. At the same time, serving queries over a KG requires the use of indexing and graph traversal methods~\cite{angles2008survey}. Further, ML methods are also instrumental to KGs. From entity matching models for entity deduplication~\cite{getoor2012entity} to link prediction models~\cite{link_prediction} for knowledge completion~\cite{lin2015learning} and natural language understanding models for fact extraction from text~\cite{miwa2016end, lin2016neural, banko2008tradeoffs}, machine learning methods have been critical to not only automate the construction of KGs~\cite{deepdive_sigmod_record} but to also enable building multi-lingual KGs~\cite{chen2016multilingual, cai2018comprehensive}.

%
%
%
%
%

%% file: sections_sigmod/conclusion.tex
\section{Conclusion}\label{sec:conclusion}
This paper described \akp, a knowledge construction and serving platform for powering entity-rich experiences across a variety of industrial use cases. We summarized the principles and design choices \akp follows to enable continuous knowledge graph construction over billions of facts and entities. We also presented deployments of \akp that support production services.

%% file: siribase_paper_sigmod.bbl

\begin{thebibliography}{88}


\ifx \showCODEN    \undefined \def \showCODEN     #1{\unskip}     \fi
\ifx \showDOI      \undefined \def \showDOI       #1{#1}\fi
\ifx \showISBNx    \undefined \def \showISBNx     #1{\unskip}     \fi
\ifx \showISBNxiii \undefined \def \showISBNxiii  #1{\unskip}     \fi
\ifx \showISSN     \undefined \def \showISSN      #1{\unskip}     \fi
\ifx \showLCCN     \undefined \def \showLCCN      #1{\unskip}     \fi
\ifx \shownote     \undefined \def \shownote      #1{#1}          \fi
\ifx \showarticletitle \undefined \def \showarticletitle #1{#1}   \fi
\ifx \showURL      \undefined \def \showURL       {\relax}        \fi
\providecommand\bibfield[2]{#2}
\providecommand\bibinfo[2]{#2}
\providecommand\natexlab[1]{#1}
\providecommand\showeprint[2][]{arXiv:#2}

\bibitem[\protect\citeauthoryear{??}{jso}{[n.d.]}]%
        {json_ld}

\newblock \bibinfo{title}{{JSON} for linking data}.
\newblock \bibinfo{howpublished}{\url{https://json-ld.org}}.
\newblock


\bibitem[\protect\citeauthoryear{Acharya, Adhikari, Agarwal, Auvray, Belgamwar,
  Biswas, Chandra, Chung, Fazel-Zarandi, Gabriel, Gao, Goel, Hakkani-Tur,
  Jezabek, Jha, Kao, Krishnan, Ku, Goyal, Lin, Liu, Mandal, Metallinou, Naik,
  Pan, Paul, Perera, Sethi, Shen, Strom, and Wang}{Acharya
  et~al\mbox{.}}{2021}]%
        {alexa_dialogue}
\bibfield{author}{\bibinfo{person}{Anish Acharya}, \bibinfo{person}{Suranjit
  Adhikari}, \bibinfo{person}{Sanchit Agarwal}, \bibinfo{person}{Vincent
  Auvray}, \bibinfo{person}{Nehal Belgamwar}, \bibinfo{person}{Arijit Biswas},
  \bibinfo{person}{Shubhra Chandra}, \bibinfo{person}{Tagyoung Chung},
  \bibinfo{person}{Maryam Fazel-Zarandi}, \bibinfo{person}{Raefer Gabriel},
  \bibinfo{person}{Shuyang Gao}, \bibinfo{person}{Rahul Goel},
  \bibinfo{person}{Dilek Hakkani-Tur}, \bibinfo{person}{Jan Jezabek},
  \bibinfo{person}{Abhay Jha}, \bibinfo{person}{Jiun-Yu Kao},
  \bibinfo{person}{Prakash Krishnan}, \bibinfo{person}{Peter Ku},
  \bibinfo{person}{Anuj Goyal}, \bibinfo{person}{Chien-Wei Lin},
  \bibinfo{person}{Qing Liu}, \bibinfo{person}{Arindam Mandal},
  \bibinfo{person}{Angeliki Metallinou}, \bibinfo{person}{Vishal Naik},
  \bibinfo{person}{Yi Pan}, \bibinfo{person}{Shachi Paul},
  \bibinfo{person}{Vittorio Perera}, \bibinfo{person}{Abhishek Sethi},
  \bibinfo{person}{Minmin Shen}, \bibinfo{person}{Nikko Strom}, {and}
  \bibinfo{person}{Eddie Wang}.} \bibinfo{year}{2021}\natexlab{}.
\newblock \showarticletitle{{A}lexa Conversations: An Extensible Data-driven
  Approach for Building Task-oriented Dialogue Systems}. In
  \bibinfo{booktitle}{\emph{Proceedings of the 2021 Conference of the North
  American Chapter of the Association for Computational Linguistics: Human
  Language Technologies: Demonstrations}}. \bibinfo{publisher}{Association for
  Computational Linguistics}, \bibinfo{address}{Online},
  \bibinfo{pages}{125--132}.
\newblock
\urldef\tempurl%
\url{https://doi.org/10.18653/v1/2021.naacl-demos.15}
\showDOI{\tempurl}


\bibitem[\protect\citeauthoryear{Aggarwal and Wang}{Aggarwal and Wang}{2010}]%
        {aggarwal2010graph}
\bibfield{author}{\bibinfo{person}{Charu~C Aggarwal} {and}
  \bibinfo{person}{Haixun Wang}.} \bibinfo{year}{2010}\natexlab{}.
\newblock \showarticletitle{Graph data management and mining: A survey of
  algorithms and applications}.
\newblock In \bibinfo{booktitle}{\emph{Managing and mining graph data}}.
  \bibinfo{publisher}{Springer}, \bibinfo{pages}{13--68}.
\newblock


\bibitem[\protect\citeauthoryear{Al~Hasan, Chaoji, Salem, and Zaki}{Al~Hasan
  et~al\mbox{.}}{[n.d.]}]%
        {link_prediction}
\bibfield{author}{\bibinfo{person}{Mohammad Al~Hasan}, \bibinfo{person}{Vineet
  Chaoji}, \bibinfo{person}{Saeed Salem}, {and} \bibinfo{person}{Mohammed
  Zaki}.}
\newblock \showarticletitle{Link prediction using supervised learning}.
\newblock


\bibitem[\protect\citeauthoryear{Angles and Gutierrez}{Angles and
  Gutierrez}{2008}]%
        {angles2008survey}
\bibfield{author}{\bibinfo{person}{Renzo Angles} {and} \bibinfo{person}{Claudio
  Gutierrez}.} \bibinfo{year}{2008}\natexlab{}.
\newblock \showarticletitle{Survey of graph database models}.
\newblock \bibinfo{journal}{\emph{ACM Computing Surveys (CSUR)}}
  \bibinfo{volume}{40}, \bibinfo{number}{1} (\bibinfo{year}{2008}),
  \bibinfo{pages}{1--39}.
\newblock


\bibitem[\protect\citeauthoryear{Banko and Etzioni}{Banko and Etzioni}{2008}]%
        {banko2008tradeoffs}
\bibfield{author}{\bibinfo{person}{Michele Banko} {and} \bibinfo{person}{Oren
  Etzioni}.} \bibinfo{year}{2008}\natexlab{}.
\newblock \showarticletitle{The tradeoffs between open and traditional relation
  extraction}. In \bibinfo{booktitle}{\emph{Proceedings of ACL-08: HLT}}.
  \bibinfo{pages}{28--36}.
\newblock


\bibitem[\protect\citeauthoryear{Bhattacharya and Getoor}{Bhattacharya and
  Getoor}{2007}]%
        {bhattacharya2007collective}
\bibfield{author}{\bibinfo{person}{Indrajit Bhattacharya} {and}
  \bibinfo{person}{Lise Getoor}.} \bibinfo{year}{2007}\natexlab{}.
\newblock \showarticletitle{Collective entity resolution in relational data}.
\newblock \bibinfo{journal}{\emph{ACM Transactions on Knowledge Discovery from
  Data (TKDD)}} \bibinfo{volume}{1}, \bibinfo{number}{1}
  (\bibinfo{year}{2007}), \bibinfo{pages}{5--es}.
\newblock


\bibitem[\protect\citeauthoryear{Bilenko and Mooney}{Bilenko and
  Mooney}{2003}]%
        {bilenko2003adaptive}
\bibfield{author}{\bibinfo{person}{Mikhail Bilenko} {and}
  \bibinfo{person}{Raymond~J Mooney}.} \bibinfo{year}{2003}\natexlab{}.
\newblock \showarticletitle{Adaptive duplicate detection using learnable string
  similarity measures}. In \bibinfo{booktitle}{\emph{Proceedings of the ninth
  ACM SIGKDD international conference on Knowledge discovery and data mining}}.
  \bibinfo{pages}{39--48}.
\newblock


\bibitem[\protect\citeauthoryear{Bollacker, Evans, Paritosh, Sturge, and
  Taylor}{Bollacker et~al\mbox{.}}{2008}]%
        {freebase}
\bibfield{author}{\bibinfo{person}{Kurt Bollacker}, \bibinfo{person}{Colin
  Evans}, \bibinfo{person}{Praveen Paritosh}, \bibinfo{person}{Tim Sturge},
  {and} \bibinfo{person}{Jamie Taylor}.} \bibinfo{year}{2008}\natexlab{}.
\newblock \showarticletitle{Freebase: a collaboratively created graph database
  for structuring human knowledge}. In \bibinfo{booktitle}{\emph{SIGMOD '08:
  Proceedings of the 2008 ACM SIGMOD international conference on Management of
  data}}. \bibinfo{publisher}{ACM}, \bibinfo{address}{New York, NY, USA},
  \bibinfo{pages}{1247--1250}.
\newblock
\urldef\tempurl%
\url{http://portal.acm.org/citation.cfm?id=1376746#}
\showURL{%
\tempurl}


\bibitem[\protect\citeauthoryear{Bordes, Usunier, Garcia-Dur\'{a}n, Weston, and
  Yakhnenko}{Bordes et~al\mbox{.}}{2013}]%
        {transe}
\bibfield{author}{\bibinfo{person}{Antoine Bordes}, \bibinfo{person}{Nicolas
  Usunier}, \bibinfo{person}{Alberto Garcia-Dur\'{a}n}, \bibinfo{person}{Jason
  Weston}, {and} \bibinfo{person}{Oksana Yakhnenko}.}
  \bibinfo{year}{2013}\natexlab{}.
\newblock \showarticletitle{Translating Embeddings for Modeling
  Multi-Relational Data}. In \bibinfo{booktitle}{\emph{Proceedings of the 26th
  International Conference on Neural Information Processing Systems - Volume
  2}} (Lake Tahoe, Nevada) \emph{(\bibinfo{series}{NIPS'13})}.
  \bibinfo{publisher}{Curran Associates Inc.}, \bibinfo{address}{Red Hook, NY,
  USA}, \bibinfo{pages}{2787–2795}.
\newblock


\bibitem[\protect\citeauthoryear{Brin and Page}{Brin and Page}{1998}]%
        {brin1994anatomy}
\bibfield{author}{\bibinfo{person}{Sergey Brin} {and} \bibinfo{person}{Lawrence
  Page}.} \bibinfo{year}{1998}\natexlab{}.
\newblock \showarticletitle{The Anatomy of a Large-Scale Hypertextual Web
  Search Engine}. In \bibinfo{booktitle}{\emph{COMPUTER NETWORKS AND ISDN
  SYSTEMS}}. \bibinfo{pages}{107--117}.
\newblock


\bibitem[\protect\citeauthoryear{Burdick, Fagin, Kolaitis, Popa, and
  Tan}{Burdick et~al\mbox{.}}{2016}]%
        {burdick2016declarative}
\bibfield{author}{\bibinfo{person}{Douglas Burdick}, \bibinfo{person}{Ronald
  Fagin}, \bibinfo{person}{Phokion~G Kolaitis}, \bibinfo{person}{Lucian Popa},
  {and} \bibinfo{person}{Wang-Chiew Tan}.} \bibinfo{year}{2016}\natexlab{}.
\newblock \showarticletitle{A declarative framework for linking entities}.
\newblock \bibinfo{journal}{\emph{ACM Transactions on Database Systems (TODS)}}
  \bibinfo{volume}{41}, \bibinfo{number}{3} (\bibinfo{year}{2016}),
  \bibinfo{pages}{1--38}.
\newblock


\bibitem[\protect\citeauthoryear{Cai, Zheng, and Chang}{Cai
  et~al\mbox{.}}{2018}]%
        {cai2018comprehensive}
\bibfield{author}{\bibinfo{person}{Hongyun Cai}, \bibinfo{person}{Vincent~W
  Zheng}, {and} \bibinfo{person}{Kevin Chen-Chuan Chang}.}
  \bibinfo{year}{2018}\natexlab{}.
\newblock \showarticletitle{A comprehensive survey of graph embedding:
  Problems, techniques, and applications}.
\newblock \bibinfo{journal}{\emph{IEEE Transactions on Knowledge and Data
  Engineering}} \bibinfo{volume}{30}, \bibinfo{number}{9}
  (\bibinfo{year}{2018}), \bibinfo{pages}{1616--1637}.
\newblock


\bibitem[\protect\citeauthoryear{Chakravarthy and Minker}{Chakravarthy and
  Minker}{1986}]%
        {chakravarthy1986multiple}
\bibfield{author}{\bibinfo{person}{Upen~S Chakravarthy} {and}
  \bibinfo{person}{Jack Minker}.} \bibinfo{year}{1986}\natexlab{}.
\newblock \showarticletitle{Multiple Query Processing in Deductive Databases
  using Query Graphs.}. In \bibinfo{booktitle}{\emph{VLDB}},
  Vol.~\bibinfo{volume}{86}. Citeseer, \bibinfo{pages}{384--391}.
\newblock


\bibitem[\protect\citeauthoryear{Chen, Tian, Yang, and Zaniolo}{Chen
  et~al\mbox{.}}{2016}]%
        {chen2016multilingual}
\bibfield{author}{\bibinfo{person}{Muhao Chen}, \bibinfo{person}{Yingtao Tian},
  \bibinfo{person}{Mohan Yang}, {and} \bibinfo{person}{Carlo Zaniolo}.}
  \bibinfo{year}{2016}\natexlab{}.
\newblock \showarticletitle{Multilingual knowledge graph embeddings for
  cross-lingual knowledge alignment}.
\newblock \bibinfo{journal}{\emph{arXiv preprint arXiv:1611.03954}}
  (\bibinfo{year}{2016}).
\newblock


\bibitem[\protect\citeauthoryear{Chen, Su, Yan, and Wang}{Chen
  et~al\mbox{.}}{2020}]%
        {chen-etal-2020-kgpt}
\bibfield{author}{\bibinfo{person}{Wenhu Chen}, \bibinfo{person}{Yu Su},
  \bibinfo{person}{Xifeng Yan}, {and} \bibinfo{person}{William~Yang Wang}.}
  \bibinfo{year}{2020}\natexlab{}.
\newblock \showarticletitle{{KGPT}: Knowledge-Grounded Pre-Training for
  Data-to-Text Generation}. In \bibinfo{booktitle}{\emph{Proceedings of the
  2020 Conference on Empirical Methods in Natural Language Processing
  (EMNLP)}}. \bibinfo{publisher}{Association for Computational Linguistics},
  \bibinfo{address}{Online}, \bibinfo{pages}{8635--8648}.
\newblock
\urldef\tempurl%
\url{https://doi.org/10.18653/v1/2020.emnlp-main.697}
\showDOI{\tempurl}


\bibitem[\protect\citeauthoryear{Cohen, Ravikumar, Fienberg,
  et~al\mbox{.}}{Cohen et~al\mbox{.}}{2003}]%
        {cohen2003comparison}
\bibfield{author}{\bibinfo{person}{William~W Cohen}, \bibinfo{person}{Pradeep
  Ravikumar}, \bibinfo{person}{Stephen~E Fienberg}, {et~al\mbox{.}}}
  \bibinfo{year}{2003}\natexlab{}.
\newblock \showarticletitle{A Comparison of String Distance Metrics for
  Name-Matching Tasks.}. In \bibinfo{booktitle}{\emph{IIWeb}},
  Vol.~\bibinfo{volume}{3}. Citeseer, \bibinfo{pages}{73--78}.
\newblock


\bibitem[\protect\citeauthoryear{Dalvi, Kumar, Pang, Ramakrishnan, Tomkins,
  Bohannon, Keerthi, and Merugu}{Dalvi et~al\mbox{.}}{2009}]%
        {webofconcepts}
\bibfield{author}{\bibinfo{person}{Nilesh Dalvi}, \bibinfo{person}{Ravi Kumar},
  \bibinfo{person}{Bo Pang}, \bibinfo{person}{Raghu Ramakrishnan},
  \bibinfo{person}{Andrew Tomkins}, \bibinfo{person}{Philip Bohannon},
  \bibinfo{person}{Sathiya Keerthi}, {and} \bibinfo{person}{Srujana Merugu}.}
  \bibinfo{year}{2009}\natexlab{}.
\newblock \showarticletitle{A Web of Concepts}. In
  \bibinfo{booktitle}{\emph{Proceedings of the Twenty-Eighth ACM
  SIGMOD-SIGACT-SIGART Symposium on Principles of Database Systems}}
  (Providence, Rhode Island, USA) \emph{(\bibinfo{series}{PODS '09})}.
  \bibinfo{pages}{1–12}.
\newblock


\bibitem[\protect\citeauthoryear{De~Sa, Ratner, R\'{e}, Shin, Wang, Wu, and
  Zhang}{De~Sa et~al\mbox{.}}{2016}]%
        {deepdive_sigmod_record}
\bibfield{author}{\bibinfo{person}{Christopher De~Sa}, \bibinfo{person}{Alex
  Ratner}, \bibinfo{person}{Christopher R\'{e}}, \bibinfo{person}{Jaeho Shin},
  \bibinfo{person}{Feiran Wang}, \bibinfo{person}{Sen Wu}, {and}
  \bibinfo{person}{Ce Zhang}.} \bibinfo{year}{2016}\natexlab{}.
\newblock \showarticletitle{DeepDive: Declarative Knowledge Base Construction}.
\newblock \bibinfo{journal}{\emph{SIGMOD Rec.}} \bibinfo{volume}{45},
  \bibinfo{number}{1} (\bibinfo{date}{June} \bibinfo{year}{2016}),
  \bibinfo{pages}{60–67}.
\newblock
\showISSN{0163-5808}


\bibitem[\protect\citeauthoryear{Doan, Halevy, and Ives}{Doan
  et~al\mbox{.}}{2012}]%
        {integration_book}
\bibfield{author}{\bibinfo{person}{AnHai Doan}, \bibinfo{person}{Alon~Y.
  Halevy}, {and} \bibinfo{person}{Zachary~G. Ives}.}
  \bibinfo{year}{2012}\natexlab{}.
\newblock \bibinfo{booktitle}{\emph{Principles of Data Integration.}}
\newblock


\bibitem[\protect\citeauthoryear{Doan, Konda, Suganthan~GC, Govind, Paulsen,
  Chandrasekhar, Martinkus, and Christie}{Doan et~al\mbox{.}}{2020}]%
        {doan2020magellan}
\bibfield{author}{\bibinfo{person}{AnHai Doan}, \bibinfo{person}{Pradap Konda},
  \bibinfo{person}{Paul Suganthan~GC}, \bibinfo{person}{Yash Govind},
  \bibinfo{person}{Derek Paulsen}, \bibinfo{person}{Kaushik Chandrasekhar},
  \bibinfo{person}{Philip Martinkus}, {and} \bibinfo{person}{Matthew
  Christie}.} \bibinfo{year}{2020}\natexlab{}.
\newblock \showarticletitle{Magellan: toward building ecosystems of entity
  matching solutions}.
\newblock \bibinfo{journal}{\emph{Commun. ACM}} \bibinfo{volume}{63},
  \bibinfo{number}{8} (\bibinfo{year}{2020}), \bibinfo{pages}{83--91}.
\newblock


\bibitem[\protect\citeauthoryear{Dong, Gabrilovich, Heitz, Horn, Lao, Murphy,
  Strohmann, Sun, and Zhang}{Dong et~al\mbox{.}}{2014}]%
        {knowledgevault}
\bibfield{author}{\bibinfo{person}{Xin Dong}, \bibinfo{person}{Evgeniy
  Gabrilovich}, \bibinfo{person}{Geremy Heitz}, \bibinfo{person}{Wilko Horn},
  \bibinfo{person}{Ni Lao}, \bibinfo{person}{Kevin Murphy},
  \bibinfo{person}{Thomas Strohmann}, \bibinfo{person}{Shaohua Sun}, {and}
  \bibinfo{person}{Wei Zhang}.} \bibinfo{year}{2014}\natexlab{}.
\newblock \showarticletitle{Knowledge vault: A web-scale approach to
  probabilistic knowledge fusion}. In \bibinfo{booktitle}{\emph{Proceedings of
  the 20th ACM SIGKDD international conference on Knowledge discovery and data
  mining}}. ACM, \bibinfo{pages}{601--610}.
\newblock
\urldef\tempurl%
\url{https://cs.cmu.edu/~nlao/publication/2014.kdd.pdf}
\showURL{%
\tempurl}


\bibitem[\protect\citeauthoryear{Dong}{Dong}{2018}]%
        {dong2018challenges}
\bibfield{author}{\bibinfo{person}{Xin~Luna Dong}.}
  \bibinfo{year}{2018}\natexlab{}.
\newblock \showarticletitle{Challenges and innovations in building a product
  knowledge graph}. In \bibinfo{booktitle}{\emph{Proceedings of the 24th ACM
  SIGKDD International Conference on Knowledge Discovery \& Data Mining}}.
  \bibinfo{pages}{2869--2869}.
\newblock


\bibitem[\protect\citeauthoryear{Dong, Berti-Equille, and Srivastava}{Dong
  et~al\mbox{.}}{2009}]%
        {luna_fusion}
\bibfield{author}{\bibinfo{person}{Xin~Luna Dong}, \bibinfo{person}{Laure
  Berti-Equille}, {and} \bibinfo{person}{Divesh Srivastava}.}
  \bibinfo{year}{2009}\natexlab{}.
\newblock \showarticletitle{Integrating Conflicting Data: The Role of Source
  Dependence}.
\newblock \bibinfo{journal}{\emph{Proc. VLDB Endow.}} \bibinfo{volume}{2},
  \bibinfo{number}{1} (\bibinfo{date}{Aug.} \bibinfo{year}{2009}),
  \bibinfo{pages}{550–561}.
\newblock
\showISSN{2150-8097}
\urldef\tempurl%
\url{https://doi.org/10.14778/1687627.1687690}
\showDOI{\tempurl}


\bibitem[\protect\citeauthoryear{Dong, Gabrilovich, Murphy, Dang, Horn,
  Lugaresi, Sun, and Zhang}{Dong et~al\mbox{.}}{2015}]%
        {knowledge_trust}
\bibfield{author}{\bibinfo{person}{Xin~Luna Dong}, \bibinfo{person}{Evgeniy
  Gabrilovich}, \bibinfo{person}{Kevin Murphy}, \bibinfo{person}{Van Dang},
  \bibinfo{person}{Wilko Horn}, \bibinfo{person}{Camillo Lugaresi},
  \bibinfo{person}{Shaohua Sun}, {and} \bibinfo{person}{Wei Zhang}.}
  \bibinfo{year}{2015}\natexlab{}.
\newblock \showarticletitle{Knowledge-Based Trust: Estimating the
  Trustworthiness of Web Sources}.
\newblock \bibinfo{journal}{\emph{Proc. VLDB Endow.}} \bibinfo{volume}{8},
  \bibinfo{number}{9} (\bibinfo{date}{May} \bibinfo{year}{2015}),
  \bibinfo{pages}{938–949}.
\newblock
\showISSN{2150-8097}
\urldef\tempurl%
\url{https://doi.org/10.14778/2777598.2777603}
\showDOI{\tempurl}


\bibitem[\protect\citeauthoryear{Dong and Naumann}{Dong and Naumann}{2009}]%
        {dong2009data}
\bibfield{author}{\bibinfo{person}{Xin~Luna Dong} {and} \bibinfo{person}{Felix
  Naumann}.} \bibinfo{year}{2009}\natexlab{}.
\newblock \showarticletitle{Data fusion: resolving data conflicts for
  integration}.
\newblock \bibinfo{journal}{\emph{Proceedings of the VLDB Endowment}}
  \bibinfo{volume}{2}, \bibinfo{number}{2} (\bibinfo{year}{2009}),
  \bibinfo{pages}{1654--1655}.
\newblock


\bibitem[\protect\citeauthoryear{Dong and Srivastava}{Dong and
  Srivastava}{2013}]%
        {dong2013big}
\bibfield{author}{\bibinfo{person}{Xin~Luna Dong} {and} \bibinfo{person}{Divesh
  Srivastava}.} \bibinfo{year}{2013}\natexlab{}.
\newblock \showarticletitle{Big data integration}. In
  \bibinfo{booktitle}{\emph{2013 IEEE 29th international conference on data
  engineering (ICDE)}}. IEEE, \bibinfo{pages}{1245--1248}.
\newblock


\bibitem[\protect\citeauthoryear{Duggan, Elmore, Stonebraker, Balazinska, Howe,
  Kepner, Madden, Maier, Mattson, and Zdonik}{Duggan et~al\mbox{.}}{2015}]%
        {stonebreaker2015polystore}
\bibfield{author}{\bibinfo{person}{Jennie Duggan}, \bibinfo{person}{Aaron~J.
  Elmore}, \bibinfo{person}{Michael Stonebraker}, \bibinfo{person}{Magda
  Balazinska}, \bibinfo{person}{Bill Howe}, \bibinfo{person}{Jeremy Kepner},
  \bibinfo{person}{Sam Madden}, \bibinfo{person}{David Maier},
  \bibinfo{person}{Tim Mattson}, {and} \bibinfo{person}{Stan Zdonik}.}
  \bibinfo{year}{2015}\natexlab{}.
\newblock \showarticletitle{The BigDAWG Polystore System}.
\newblock \bibinfo{journal}{\emph{SIGMOD Rec.}} \bibinfo{volume}{44},
  \bibinfo{number}{2} (\bibinfo{date}{Aug.} \bibinfo{year}{2015}),
  \bibinfo{pages}{11–16}.
\newblock
\showISSN{0163-5808}
\urldef\tempurl%
\url{https://doi.org/10.1145/2814710.2814713}
\showDOI{\tempurl}


\bibitem[\protect\citeauthoryear{Elmagarmid, Ilyas, Ouzzani, Quian{\'e}-Ruiz,
  Tang, and Yin}{Elmagarmid et~al\mbox{.}}{2014}]%
        {elmagarmid2014nadeef}
\bibfield{author}{\bibinfo{person}{Ahmed Elmagarmid}, \bibinfo{person}{Ihab~F
  Ilyas}, \bibinfo{person}{Mourad Ouzzani}, \bibinfo{person}{Jorge-Arnulfo
  Quian{\'e}-Ruiz}, \bibinfo{person}{Nan Tang}, {and} \bibinfo{person}{Si
  Yin}.} \bibinfo{year}{2014}\natexlab{}.
\newblock \showarticletitle{NADEEF/ER: Generic and interactive entity
  resolution}. In \bibinfo{booktitle}{\emph{Proceedings of the 2014 ACM SIGMOD
  international conference on Management of data}}.
  \bibinfo{pages}{1071--1074}.
\newblock


\bibitem[\protect\citeauthoryear{Etzioni, Cafarella, Downey, Kok, Popescu,
  Shaked, Soderland, Weld, and Yates}{Etzioni et~al\mbox{.}}{2004}]%
        {knowitall}
\bibfield{author}{\bibinfo{person}{Oren Etzioni}, \bibinfo{person}{Michael
  Cafarella}, \bibinfo{person}{Doug Downey}, \bibinfo{person}{Stanley Kok},
  \bibinfo{person}{Ana-Maria Popescu}, \bibinfo{person}{Tal Shaked},
  \bibinfo{person}{Stephen Soderland}, \bibinfo{person}{Daniel~S. Weld}, {and}
  \bibinfo{person}{Alexander Yates}.} \bibinfo{year}{2004}\natexlab{}.
\newblock \showarticletitle{Web-Scale Information Extraction in Knowitall:
  (Preliminary Results)}. In \bibinfo{booktitle}{\emph{Proceedings of the 13th
  International Conference on World Wide Web}} (New York, NY, USA)
  \emph{(\bibinfo{series}{WWW '04})}. \bibinfo{pages}{100–110}.
\newblock


\bibitem[\protect\citeauthoryear{Fan, Raj, and Patel}{Fan
  et~al\mbox{.}}{2015}]%
        {fan2015case}
\bibfield{author}{\bibinfo{person}{Jing Fan}, \bibinfo{person}{Adalbert
  Gerald~Soosai Raj}, {and} \bibinfo{person}{Jignesh~M Patel}.}
  \bibinfo{year}{2015}\natexlab{}.
\newblock \showarticletitle{The Case Against Specialized Graph Analytics
  Engines.}. In \bibinfo{booktitle}{\emph{CIDR}}.
\newblock


\bibitem[\protect\citeauthoryear{Fast, Chen, Mendelsohn, Bassen, and
  Bernstein}{Fast et~al\mbox{.}}{2018}]%
        {fast2018iris}
\bibfield{author}{\bibinfo{person}{Ethan Fast}, \bibinfo{person}{Binbin Chen},
  \bibinfo{person}{Julia Mendelsohn}, \bibinfo{person}{Jonathan Bassen}, {and}
  \bibinfo{person}{Michael~S Bernstein}.} \bibinfo{year}{2018}\natexlab{}.
\newblock \showarticletitle{Iris: A conversational agent for complex tasks}. In
  \bibinfo{booktitle}{\emph{Proceedings of the 2018 CHI Conference on Human
  Factors in Computing Systems}}. \bibinfo{pages}{1--12}.
\newblock


\bibitem[\protect\citeauthoryear{Gao, Liang, Han, Yakout, and Mohamed}{Gao
  et~al\mbox{.}}{2018}]%
        {satori}
\bibfield{author}{\bibinfo{person}{Yuqing Gao}, \bibinfo{person}{Jisheng
  Liang}, \bibinfo{person}{Benjamin Han}, \bibinfo{person}{Mohamed Yakout},
  {and} \bibinfo{person}{Ahmed Mohamed}.} \bibinfo{year}{2018}\natexlab{}.
\newblock \showarticletitle{Building a large-scale, accurate and fresh
  knowledge graph}.
\newblock \bibinfo{journal}{\emph{KDD-2018, Tutorial}}  \bibinfo{volume}{39}
  (\bibinfo{year}{2018}), \bibinfo{pages}{1939--1374}.
\newblock


\bibitem[\protect\citeauthoryear{Geifman and El-Yaniv}{Geifman and
  El-Yaniv}{2017}]%
        {rejection}
\bibfield{author}{\bibinfo{person}{Yonatan Geifman} {and} \bibinfo{person}{Ran
  El-Yaniv}.} \bibinfo{year}{2017}\natexlab{}.
\newblock \showarticletitle{Selective Classification for Deep Neural Networks}.
  In \bibinfo{booktitle}{\emph{Proceedings of the 31st International Conference
  on Neural Information Processing Systems}} (Long Beach, California, USA)
  \emph{(\bibinfo{series}{NIPS'17})}. \bibinfo{publisher}{Curran Associates
  Inc.}, \bibinfo{address}{Red Hook, NY, USA}, \bibinfo{pages}{4885–4894}.
\newblock
\showISBNx{9781510860964}


\bibitem[\protect\citeauthoryear{Getoor and Machanavajjhala}{Getoor and
  Machanavajjhala}{2012}]%
        {getoor2012entity}
\bibfield{author}{\bibinfo{person}{Lise Getoor} {and} \bibinfo{person}{Ashwin
  Machanavajjhala}.} \bibinfo{year}{2012}\natexlab{}.
\newblock \showarticletitle{Entity resolution: theory, practice \& open
  challenges}.
\newblock \bibinfo{journal}{\emph{Proceedings of the VLDB Endowment}}
  \bibinfo{volume}{5}, \bibinfo{number}{12} (\bibinfo{year}{2012}),
  \bibinfo{pages}{2018--2019}.
\newblock


\bibitem[\protect\citeauthoryear{Grishman and Sundheim}{Grishman and
  Sundheim}{1996}]%
        {grishman-sundheim-1996-message}
\bibfield{author}{\bibinfo{person}{Ralph Grishman} {and} \bibinfo{person}{Beth
  Sundheim}.} \bibinfo{year}{1996}\natexlab{}.
\newblock \showarticletitle{{M}essage {U}nderstanding {C}onference- 6: A Brief
  History}. In \bibinfo{booktitle}{\emph{{COLING} 1996 Volume 1: The 16th
  International Conference on Computational Linguistics}}.
\newblock
\urldef\tempurl%
\url{https://aclanthology.org/C96-1079}
\showURL{%
\tempurl}


\bibitem[\protect\citeauthoryear{Gruenheid, Dong, and Srivastava}{Gruenheid
  et~al\mbox{.}}{2014}]%
        {gruenheid2014incremental}
\bibfield{author}{\bibinfo{person}{Anja Gruenheid}, \bibinfo{person}{Xin~Luna
  Dong}, {and} \bibinfo{person}{Divesh Srivastava}.}
  \bibinfo{year}{2014}\natexlab{}.
\newblock \showarticletitle{Incremental record linkage}.
\newblock \bibinfo{journal}{\emph{Proceedings of the VLDB Endowment}}
  \bibinfo{volume}{7}, \bibinfo{number}{9} (\bibinfo{year}{2014}),
  \bibinfo{pages}{697--708}.
\newblock


\bibitem[\protect\citeauthoryear{Heidari, McGrath, Ilyas, and
  Rekatsinas}{Heidari et~al\mbox{.}}{2019}]%
        {heidari2019holodetect}
\bibfield{author}{\bibinfo{person}{Alireza Heidari}, \bibinfo{person}{Joshua
  McGrath}, \bibinfo{person}{Ihab~F Ilyas}, {and} \bibinfo{person}{Theodoros
  Rekatsinas}.} \bibinfo{year}{2019}\natexlab{}.
\newblock \showarticletitle{Holodetect: Few-shot learning for error detection}.
  In \bibinfo{booktitle}{\emph{Proceedings of the 2019 International Conference
  on Management of Data}}. \bibinfo{pages}{829--846}.
\newblock


\bibitem[\protect\citeauthoryear{Heidari, Michalopoulos, Kushagra, Ilyas, and
  Rekatsinas}{Heidari et~al\mbox{.}}{2020}]%
        {heidari2020record}
\bibfield{author}{\bibinfo{person}{Alireza Heidari}, \bibinfo{person}{George
  Michalopoulos}, \bibinfo{person}{Shrinu Kushagra}, \bibinfo{person}{Ihab~F
  Ilyas}, {and} \bibinfo{person}{Theodoros Rekatsinas}.}
  \bibinfo{year}{2020}\natexlab{}.
\newblock \showarticletitle{Record fusion: A learning approach}.
\newblock \bibinfo{journal}{\emph{arXiv preprint arXiv:2006.10208}}
  (\bibinfo{year}{2020}).
\newblock


\bibitem[\protect\citeauthoryear{Ilyas and Chu}{Ilyas and Chu}{2019}]%
        {ilyas2019data}
\bibfield{author}{\bibinfo{person}{Ihab~F Ilyas} {and} \bibinfo{person}{Xu
  Chu}.} \bibinfo{year}{2019}\natexlab{}.
\newblock \bibinfo{booktitle}{\emph{Data cleaning}}.
\newblock \bibinfo{publisher}{Morgan \& Claypool}.
\newblock


\bibitem[\protect\citeauthoryear{Jindal, Rawlani, Wu, Madden, Deshpande, and
  Stonebraker}{Jindal et~al\mbox{.}}{2014}]%
        {jindal2014vertexica}
\bibfield{author}{\bibinfo{person}{Alekh Jindal}, \bibinfo{person}{Praynaa
  Rawlani}, \bibinfo{person}{Eugene Wu}, \bibinfo{person}{Samuel Madden},
  \bibinfo{person}{Amol Deshpande}, {and} \bibinfo{person}{Mike Stonebraker}.}
  \bibinfo{year}{2014}\natexlab{}.
\newblock \showarticletitle{Vertexica: your relational friend for graph
  analytics!}
\newblock  (\bibinfo{year}{2014}).
\newblock


\bibitem[\protect\citeauthoryear{John, Potti, and Patel}{John
  et~al\mbox{.}}{2017}]%
        {john2017ava}
\bibfield{author}{\bibinfo{person}{Rogers Jeffrey~Leo John},
  \bibinfo{person}{Navneet Potti}, {and} \bibinfo{person}{Jignesh~M Patel}.}
  \bibinfo{year}{2017}\natexlab{}.
\newblock \showarticletitle{Ava: From Data to Insights Through Conversations.}.
  In \bibinfo{booktitle}{\emph{CIDR}}.
\newblock


\bibitem[\protect\citeauthoryear{Kim, Jernite, Sontag, and Rush}{Kim
  et~al\mbox{.}}{2016}]%
        {kim2016character}
\bibfield{author}{\bibinfo{person}{Yoon Kim}, \bibinfo{person}{Yacine Jernite},
  \bibinfo{person}{David Sontag}, {and} \bibinfo{person}{Alexander~M Rush}.}
  \bibinfo{year}{2016}\natexlab{}.
\newblock \showarticletitle{Character-aware neural language models}. In
  \bibinfo{booktitle}{\emph{Thirtieth AAAI conference on artificial
  intelligence}}.
\newblock


\bibitem[\protect\citeauthoryear{Koudas, Sarawagi, and Srivastava}{Koudas
  et~al\mbox{.}}{2006}]%
        {koudas2006record}
\bibfield{author}{\bibinfo{person}{Nick Koudas}, \bibinfo{person}{Sunita
  Sarawagi}, {and} \bibinfo{person}{Divesh Srivastava}.}
  \bibinfo{year}{2006}\natexlab{}.
\newblock \showarticletitle{Record linkage: similarity measures and
  algorithms}. In \bibinfo{booktitle}{\emph{Proceedings of the 2006 ACM SIGMOD
  international conference on Management of data}}. \bibinfo{pages}{802--803}.
\newblock


\bibitem[\protect\citeauthoryear{Lample, Ballesteros, Subramanian, Kawakami,
  and Dyer}{Lample et~al\mbox{.}}{2016}]%
        {lample2016neural}
\bibfield{author}{\bibinfo{person}{Guillaume Lample}, \bibinfo{person}{Miguel
  Ballesteros}, \bibinfo{person}{Sandeep Subramanian}, \bibinfo{person}{Kazuya
  Kawakami}, {and} \bibinfo{person}{Chris Dyer}.}
  \bibinfo{year}{2016}\natexlab{}.
\newblock \showarticletitle{Neural architectures for named entity recognition}.
\newblock \bibinfo{journal}{\emph{arXiv preprint arXiv:1603.01360}}
  (\bibinfo{year}{2016}).
\newblock


\bibitem[\protect\citeauthoryear{Lassila, Swick, et~al\mbox{.}}{Lassila
  et~al\mbox{.}}{1998}]%
        {rdf}
\bibfield{author}{\bibinfo{person}{Ora Lassila}, \bibinfo{person}{Ralph~R
  Swick}, {et~al\mbox{.}}} \bibinfo{year}{1998}\natexlab{}.
\newblock \showarticletitle{Resource description framework (RDF) model and
  syntax specification}.
\newblock  (\bibinfo{year}{1998}).
\newblock


\bibitem[\protect\citeauthoryear{Lehmann, Isele, Jakob, Jentzsch, Kontokostas,
  Mendes, Hellmann, Morsey, van Kleef, Auer, and Bizer}{Lehmann
  et~al\mbox{.}}{2015}]%
        {dbpedia}
\bibfield{author}{\bibinfo{person}{Jens Lehmann}, \bibinfo{person}{Robert
  Isele}, \bibinfo{person}{Max Jakob}, \bibinfo{person}{Anja Jentzsch},
  \bibinfo{person}{Dimitris Kontokostas}, \bibinfo{person}{Pablo~N. Mendes},
  \bibinfo{person}{Sebastian Hellmann}, \bibinfo{person}{Mohamed Morsey},
  \bibinfo{person}{Patrick van Kleef}, \bibinfo{person}{Sören Auer}, {and}
  \bibinfo{person}{Christian Bizer}.} \bibinfo{year}{2015}\natexlab{}.
\newblock \showarticletitle{DBpedia - A large-scale, multilingual knowledge
  base extracted from Wikipedia.}
\newblock   \bibinfo{volume}{6} (\bibinfo{year}{2015}).
\newblock


\bibitem[\protect\citeauthoryear{Lenzerini}{Lenzerini}{2002}]%
        {lenzerini2002data}
\bibfield{author}{\bibinfo{person}{Maurizio Lenzerini}.}
  \bibinfo{year}{2002}\natexlab{}.
\newblock \showarticletitle{Data integration: A theoretical perspective}. In
  \bibinfo{booktitle}{\emph{Proceedings of the twenty-first ACM
  SIGMOD-SIGACT-SIGART symposium on Principles of database systems}}.
  \bibinfo{pages}{233--246}.
\newblock


\bibitem[\protect\citeauthoryear{Lerer, Wu, Shen, Lacroix, Wehrstedt, Bose, and
  Peysakhovich}{Lerer et~al\mbox{.}}{2019}]%
        {biggraph}
\bibfield{author}{\bibinfo{person}{Adam Lerer}, \bibinfo{person}{Ledell Wu},
  \bibinfo{person}{Jiajun Shen}, \bibinfo{person}{Timoth{\'{e}}e Lacroix},
  \bibinfo{person}{Luca Wehrstedt}, \bibinfo{person}{Abhijit Bose}, {and}
  \bibinfo{person}{Alexander Peysakhovich}.} \bibinfo{year}{2019}\natexlab{}.
\newblock \showarticletitle{PyTorch-BigGraph: {A} Large-scale Graph Embedding
  System}.
\newblock \bibinfo{journal}{\emph{CoRR}}  \bibinfo{volume}{abs/1903.12287}
  (\bibinfo{year}{2019}).
\newblock
\showeprint[arXiv]{1903.12287}
\urldef\tempurl%
\url{http://arxiv.org/abs/1903.12287}
\showURL{%
\tempurl}


\bibitem[\protect\citeauthoryear{Li, Gao, Meng, Li, Su, Zhao, Fan, and Han}{Li
  et~al\mbox{.}}{2016}]%
        {li2016survey}
\bibfield{author}{\bibinfo{person}{Yaliang Li}, \bibinfo{person}{Jing Gao},
  \bibinfo{person}{Chuishi Meng}, \bibinfo{person}{Qi Li}, \bibinfo{person}{Lu
  Su}, \bibinfo{person}{Bo Zhao}, \bibinfo{person}{Wei Fan}, {and}
  \bibinfo{person}{Jiawei Han}.} \bibinfo{year}{2016}\natexlab{}.
\newblock \showarticletitle{A survey on truth discovery}.
\newblock \bibinfo{journal}{\emph{ACM Sigkdd Explorations Newsletter}}
  \bibinfo{volume}{17}, \bibinfo{number}{2} (\bibinfo{year}{2016}),
  \bibinfo{pages}{1--16}.
\newblock


\bibitem[\protect\citeauthoryear{Lin, Liu, Sun, Liu, and Zhu}{Lin
  et~al\mbox{.}}{2015}]%
        {lin2015learning}
\bibfield{author}{\bibinfo{person}{Yankai Lin}, \bibinfo{person}{Zhiyuan Liu},
  \bibinfo{person}{Maosong Sun}, \bibinfo{person}{Yang Liu}, {and}
  \bibinfo{person}{Xuan Zhu}.} \bibinfo{year}{2015}\natexlab{}.
\newblock \showarticletitle{Learning entity and relation embeddings for
  knowledge graph completion}. In \bibinfo{booktitle}{\emph{Twenty-ninth AAAI
  conference on artificial intelligence}}.
\newblock


\bibitem[\protect\citeauthoryear{Lin, Shen, Liu, Luan, and Sun}{Lin
  et~al\mbox{.}}{2016}]%
        {lin2016neural}
\bibfield{author}{\bibinfo{person}{Yankai Lin}, \bibinfo{person}{Shiqi Shen},
  \bibinfo{person}{Zhiyuan Liu}, \bibinfo{person}{Huanbo Luan}, {and}
  \bibinfo{person}{Maosong Sun}.} \bibinfo{year}{2016}\natexlab{}.
\newblock \showarticletitle{Neural relation extraction with selective attention
  over instances}. In \bibinfo{booktitle}{\emph{Proceedings of the 54th Annual
  Meeting of the Association for Computational Linguistics (Volume 1: Long
  Papers)}}. \bibinfo{pages}{2124--2133}.
\newblock


\bibitem[\protect\citeauthoryear{Mintz, Bills, Snow, and Jurafsky}{Mintz
  et~al\mbox{.}}{2009}]%
        {mintz2009distant}
\bibfield{author}{\bibinfo{person}{Mike Mintz}, \bibinfo{person}{Steven Bills},
  \bibinfo{person}{Rion Snow}, {and} \bibinfo{person}{Dan Jurafsky}.}
  \bibinfo{year}{2009}\natexlab{}.
\newblock \showarticletitle{Distant supervision for relation extraction without
  labeled data}. In \bibinfo{booktitle}{\emph{Proceedings of the Joint
  Conference of the 47th Annual Meeting of the ACL and the 4th International
  Joint Conference on Natural Language Processing of the AFNLP}}.
  \bibinfo{pages}{1003--1011}.
\newblock


\bibitem[\protect\citeauthoryear{Mitchell, Cohen, Hruscha, Talukdar,
  Betteridge, Carlson, Dalvi, Gardner, Kisiel, Krishnamurthy, Lao, Mazaitis,
  Mohammad, Nakashole, Platanios, Ritter, Samadi, Settles, Wang, Wijaya, Gupta,
  Chen, Saparov, Greaves, and Welling}{Mitchell et~al\mbox{.}}{2015}]%
        {nell}
\bibfield{author}{\bibinfo{person}{T. Mitchell}, \bibinfo{person}{W. Cohen},
  \bibinfo{person}{E. Hruscha}, \bibinfo{person}{P. Talukdar},
  \bibinfo{person}{J. Betteridge}, \bibinfo{person}{A. Carlson},
  \bibinfo{person}{B. Dalvi}, \bibinfo{person}{M. Gardner}, \bibinfo{person}{B.
  Kisiel}, \bibinfo{person}{J. Krishnamurthy}, \bibinfo{person}{N. Lao},
  \bibinfo{person}{K. Mazaitis}, \bibinfo{person}{T. Mohammad},
  \bibinfo{person}{N. Nakashole}, \bibinfo{person}{E. Platanios},
  \bibinfo{person}{A. Ritter}, \bibinfo{person}{M. Samadi}, \bibinfo{person}{B.
  Settles}, \bibinfo{person}{R. Wang}, \bibinfo{person}{D. Wijaya},
  \bibinfo{person}{A. Gupta}, \bibinfo{person}{X. Chen}, \bibinfo{person}{A.
  Saparov}, \bibinfo{person}{M. Greaves}, {and} \bibinfo{person}{J. Welling}.}
  \bibinfo{year}{2015}\natexlab{}.
\newblock \showarticletitle{Never-Ending Learning}. In
  \bibinfo{booktitle}{\emph{AAAI}}.
\newblock
\urldef\tempurl%
\url{http://www.cs.cmu.edu/~wcohen/pubs.html}
\showURL{%
\tempurl}
\newblock
\shownote{: Never-Ending Learning in AAAI-2015.}


\bibitem[\protect\citeauthoryear{Miwa and Bansal}{Miwa and Bansal}{2016}]%
        {miwa2016end}
\bibfield{author}{\bibinfo{person}{Makoto Miwa} {and} \bibinfo{person}{Mohit
  Bansal}.} \bibinfo{year}{2016}\natexlab{}.
\newblock \showarticletitle{End-to-end relation extraction using lstms on
  sequences and tree structures}.
\newblock \bibinfo{journal}{\emph{arXiv preprint arXiv:1601.00770}}
  (\bibinfo{year}{2016}).
\newblock


\bibitem[\protect\citeauthoryear{Mohoney, Waleffe, Xu, Rekatsinas, and
  Venkataraman}{Mohoney et~al\mbox{.}}{2021}]%
        {marius}
\bibfield{author}{\bibinfo{person}{Jason Mohoney}, \bibinfo{person}{Roger
  Waleffe}, \bibinfo{person}{Henry Xu}, \bibinfo{person}{Theodoros Rekatsinas},
  {and} \bibinfo{person}{Shivaram Venkataraman}.}
  \bibinfo{year}{2021}\natexlab{}.
\newblock \showarticletitle{Marius: Learning Massive Graph Embeddings on a
  Single Machine}. In \bibinfo{booktitle}{\emph{15th {USENIX} Symposium on
  Operating Systems Design and Implementation, {OSDI} 2021, July 14-16, 2021}},
  \bibfield{editor}{\bibinfo{person}{Angela~Demke Brown} {and}
  \bibinfo{person}{Jay~R. Lorch}} (Eds.). \bibinfo{publisher}{{USENIX}
  Association}, \bibinfo{pages}{533--549}.
\newblock


\bibitem[\protect\citeauthoryear{Mudgal, Li, Rekatsinas, Doan, Park, Krishnan,
  Deep, Arcaute, and Raghavendra}{Mudgal et~al\mbox{.}}{2018}]%
        {mudgal2018deep}
\bibfield{author}{\bibinfo{person}{Sidharth Mudgal}, \bibinfo{person}{Han Li},
  \bibinfo{person}{Theodoros Rekatsinas}, \bibinfo{person}{AnHai Doan},
  \bibinfo{person}{Youngchoon Park}, \bibinfo{person}{Ganesh Krishnan},
  \bibinfo{person}{Rohit Deep}, \bibinfo{person}{Esteban Arcaute}, {and}
  \bibinfo{person}{Vijay Raghavendra}.} \bibinfo{year}{2018}\natexlab{}.
\newblock \showarticletitle{Deep learning for entity matching: A design space
  exploration}. In \bibinfo{booktitle}{\emph{Proceedings of the 2018
  International Conference on Management of Data}}. \bibinfo{pages}{19--34}.
\newblock


\bibitem[\protect\citeauthoryear{Mulang', Singh, Prabhu, Nadgeri, Hoffart, and
  Lehmann}{Mulang' et~al\mbox{.}}{2020}]%
        {mulang}
\bibfield{author}{\bibinfo{person}{Isaiah~Onando Mulang'},
  \bibinfo{person}{Kuldeep Singh}, \bibinfo{person}{Chaitali Prabhu},
  \bibinfo{person}{Abhishek Nadgeri}, \bibinfo{person}{Johannes Hoffart}, {and}
  \bibinfo{person}{Jens Lehmann}.} \bibinfo{year}{2020}\natexlab{}.
\newblock \showarticletitle{Evaluating the Impact of Knowledge Graph Context on
  Entity Disambiguation Models}. In \bibinfo{booktitle}{\emph{Proceedings of
  the 29th ACM International Conference on Information and Knowledge
  Management}} (Virtual Event, Ireland) \emph{(\bibinfo{series}{CIKM '20})}.
  \bibinfo{publisher}{Association for Computing Machinery},
  \bibinfo{address}{New York, NY, USA}, \bibinfo{pages}{2157–2160}.
\newblock
\showISBNx{9781450368599}
\urldef\tempurl%
\url{https://doi.org/10.1145/3340531.3412159}
\showDOI{\tempurl}


\bibitem[\protect\citeauthoryear{Nadeau and Sekine}{Nadeau and Sekine}{2007}]%
        {nadeau2007survey}
\bibfield{author}{\bibinfo{person}{David Nadeau} {and} \bibinfo{person}{Satoshi
  Sekine}.} \bibinfo{year}{2007}\natexlab{}.
\newblock \showarticletitle{A survey of named entity recognition and
  classification}.
\newblock \bibinfo{journal}{\emph{Lingvisticae Investigationes}}
  \bibinfo{volume}{30}, \bibinfo{number}{1} (\bibinfo{year}{2007}),
  \bibinfo{pages}{3--26}.
\newblock


\bibitem[\protect\citeauthoryear{Nguyen, Hoffart, Theobald, and Weikum}{Nguyen
  et~al\mbox{.}}{2014}]%
        {nguyen2014aida}
\bibfield{author}{\bibinfo{person}{Dat~Ba Nguyen}, \bibinfo{person}{Johannes
  Hoffart}, \bibinfo{person}{Martin Theobald}, {and} \bibinfo{person}{Gerhard
  Weikum}.} \bibinfo{year}{2014}\natexlab{}.
\newblock \showarticletitle{Aida-light: High-throughput named-entity
  disambiguation}. In \bibinfo{booktitle}{\emph{LDOW}}.
\newblock


\bibitem[\protect\citeauthoryear{Noy, Gao, Jain, Narayanan, Patterson, and
  Taylor}{Noy et~al\mbox{.}}{2019}]%
        {industry_kgs}
\bibfield{author}{\bibinfo{person}{Natasha Noy}, \bibinfo{person}{Yuqing Gao},
  \bibinfo{person}{Anshu Jain}, \bibinfo{person}{Anant Narayanan},
  \bibinfo{person}{Alan Patterson}, {and} \bibinfo{person}{Jamie Taylor}.}
  \bibinfo{year}{2019}\natexlab{}.
\newblock \showarticletitle{Industry-Scale Knowledge Graphs: Lessons and
  Challenges}.
\newblock \bibinfo{journal}{\emph{Commun. ACM}} \bibinfo{volume}{62},
  \bibinfo{number}{8} (\bibinfo{date}{July} \bibinfo{year}{2019}),
  \bibinfo{pages}{36–43}.
\newblock
\showISSN{0001-0782}
\urldef\tempurl%
\url{https://doi.org/10.1145/3331166}
\showDOI{\tempurl}


\bibitem[\protect\citeauthoryear{Orr, Leszczynski, Arora, Wu, Guha, Ling, and
  Re}{Orr et~al\mbox{.}}{2021}]%
        {orr2020bootleg}
\bibfield{author}{\bibinfo{person}{Laurel Orr}, \bibinfo{person}{Megan
  Leszczynski}, \bibinfo{person}{Simran Arora}, \bibinfo{person}{Sen Wu},
  \bibinfo{person}{Neel Guha}, \bibinfo{person}{Xiao Ling}, {and}
  \bibinfo{person}{Christopher Re}.} \bibinfo{year}{2021}\natexlab{}.
\newblock \showarticletitle{Bootleg: Chasing the tail with self-supervised
  named entity disambiguation}.
\newblock \bibinfo{journal}{\emph{CIDR}} (\bibinfo{year}{2021}).
\newblock


\bibitem[\protect\citeauthoryear{Pan, Papailiopoulos, Oymak, Recht,
  Ramchandran, and Jordan}{Pan et~al\mbox{.}}{2015}]%
        {parallel_correlation_clustering}
\bibfield{author}{\bibinfo{person}{Xinghao Pan}, \bibinfo{person}{Dimitris
  Papailiopoulos}, \bibinfo{person}{Samet Oymak}, \bibinfo{person}{Benjamin
  Recht}, \bibinfo{person}{Kannan Ramchandran}, {and}
  \bibinfo{person}{Michael~I. Jordan}.} \bibinfo{year}{2015}\natexlab{}.
\newblock \showarticletitle{Parallel Correlation Clustering on Big Graphs}. In
  \bibinfo{booktitle}{\emph{Proceedings of the 28th International Conference on
  Neural Information Processing Systems - Volume 1}} (Montreal, Canada)
  \emph{(\bibinfo{series}{NIPS'15})}. \bibinfo{publisher}{MIT Press},
  \bibinfo{address}{Cambridge, MA, USA}, \bibinfo{pages}{82–90}.
\newblock


\bibitem[\protect\citeauthoryear{Papadakis, Skoutas, Thanos, and
  Palpanas}{Papadakis et~al\mbox{.}}{2020}]%
        {papadakis2020blocking}
\bibfield{author}{\bibinfo{person}{George Papadakis},
  \bibinfo{person}{Dimitrios Skoutas}, \bibinfo{person}{Emmanouil Thanos},
  {and} \bibinfo{person}{Themis Palpanas}.} \bibinfo{year}{2020}\natexlab{}.
\newblock \showarticletitle{Blocking and filtering techniques for entity
  resolution: A survey}.
\newblock \bibinfo{journal}{\emph{ACM Computing Surveys (CSUR)}}
  \bibinfo{volume}{53}, \bibinfo{number}{2} (\bibinfo{year}{2020}),
  \bibinfo{pages}{1--42}.
\newblock


\bibitem[\protect\citeauthoryear{Rastogi, Dalvi, and Garofalakis}{Rastogi
  et~al\mbox{.}}{2011}]%
        {collective_entity_matching}
\bibfield{author}{\bibinfo{person}{Vibhor Rastogi}, \bibinfo{person}{Nilesh
  Dalvi}, {and} \bibinfo{person}{Minos Garofalakis}.}
  \bibinfo{year}{2011}\natexlab{}.
\newblock \showarticletitle{Large-Scale Collective Entity Matching}.
\newblock \bibinfo{journal}{\emph{Proc. VLDB Endow.}} \bibinfo{volume}{4},
  \bibinfo{number}{4} (\bibinfo{date}{Jan.} \bibinfo{year}{2011}),
  \bibinfo{pages}{208–218}.
\newblock
\showISSN{2150-8097}
\urldef\tempurl%
\url{https://doi.org/10.14778/1938545.1938546}
\showDOI{\tempurl}


\bibitem[\protect\citeauthoryear{Rekatsinas, Chu, Ilyas, and R\'{e}}{Rekatsinas
  et~al\mbox{.}}{2017a}]%
        {holoclean}
\bibfield{author}{\bibinfo{person}{Theodoros Rekatsinas}, \bibinfo{person}{Xu
  Chu}, \bibinfo{person}{Ihab~F. Ilyas}, {and} \bibinfo{person}{Christopher
  R\'{e}}.} \bibinfo{year}{2017}\natexlab{a}.
\newblock \showarticletitle{HoloClean: Holistic Data Repairs with Probabilistic
  Inference}.
\newblock \bibinfo{journal}{\emph{Proc. VLDB Endow.}} \bibinfo{volume}{10},
  \bibinfo{number}{11} (\bibinfo{date}{Aug.} \bibinfo{year}{2017}),
  \bibinfo{pages}{1190–1201}.
\newblock
\showISSN{2150-8097}
\urldef\tempurl%
\url{https://doi.org/10.14778/3137628.3137631}
\showDOI{\tempurl}


\bibitem[\protect\citeauthoryear{Rekatsinas, Joglekar, Garcia-Molina,
  Parameswaran, and R\'{e}}{Rekatsinas et~al\mbox{.}}{2017b}]%
        {slimfast}
\bibfield{author}{\bibinfo{person}{Theodoros Rekatsinas},
  \bibinfo{person}{Manas Joglekar}, \bibinfo{person}{Hector Garcia-Molina},
  \bibinfo{person}{Aditya Parameswaran}, {and} \bibinfo{person}{Christopher
  R\'{e}}.} \bibinfo{year}{2017}\natexlab{b}.
\newblock \showarticletitle{SLiMFast: Guaranteed Results for Data Fusion and
  Source Reliability}. In \bibinfo{booktitle}{\emph{Proceedings of the 2017 ACM
  International Conference on Management of Data}}
  \emph{(\bibinfo{series}{SIGMOD '17})}. \bibinfo{publisher}{Association for
  Computing Machinery}, \bibinfo{address}{New York, NY, USA},
  \bibinfo{pages}{1399–1414}.
\newblock
\showISBNx{9781450341974}
\urldef\tempurl%
\url{https://doi.org/10.1145/3035918.3035951}
\showDOI{\tempurl}


\bibitem[\protect\citeauthoryear{Saeedi, Peukert, and Rahm}{Saeedi
  et~al\mbox{.}}{2017}]%
        {saeedi2017comparative}
\bibfield{author}{\bibinfo{person}{Alieh Saeedi}, \bibinfo{person}{Eric
  Peukert}, {and} \bibinfo{person}{Erhard Rahm}.}
  \bibinfo{year}{2017}\natexlab{}.
\newblock \showarticletitle{Comparative evaluation of distributed clustering
  schemes for multi-source entity resolution}. In
  \bibinfo{booktitle}{\emph{European Conference on Advances in Databases and
  Information Systems}}. Springer, \bibinfo{pages}{278--293}.
\newblock


\bibitem[\protect\citeauthoryear{Salehpour and Davis}{Salehpour and
  Davis}{2020}]%
        {extended_triples}
\bibfield{author}{\bibinfo{person}{Masoud Salehpour} {and}
  \bibinfo{person}{Joseph~G. Davis}.} \bibinfo{year}{2020}\natexlab{}.
\newblock \showarticletitle{The Effects of Different {JSON} Representations on
  Querying Knowledge Graphs}.
\newblock \bibinfo{journal}{\emph{CoRR}}  \bibinfo{volume}{abs/2004.04286}
  (\bibinfo{year}{2020}).
\newblock
\showeprint[arXiv]{2004.04286}
\urldef\tempurl%
\url{https://arxiv.org/abs/2004.04286}
\showURL{%
\tempurl}


\bibitem[\protect\citeauthoryear{Sellis}{Sellis}{1986}]%
        {sellis86global}
\bibfield{author}{\bibinfo{person}{Timos~K. Sellis}.}
  \bibinfo{year}{1986}\natexlab{}.
\newblock \showarticletitle{Global Query Optimization}. In
  \bibinfo{booktitle}{\emph{Proceedings of the 1986 ACM SIGMOD International
  Conference on Management of Data}} (Washington, D.C., USA)
  \emph{(\bibinfo{series}{SIGMOD '86})}. \bibinfo{publisher}{Association for
  Computing Machinery}, \bibinfo{address}{New York, NY, USA},
  \bibinfo{pages}{191–205}.
\newblock
\showISBNx{0897911911}
\urldef\tempurl%
\url{https://doi.org/10.1145/16894.16874}
\showDOI{\tempurl}


\bibitem[\protect\citeauthoryear{Singh, Meduri, Elmagarmid, Madden, Papotti,
  Quian{\'e}-Ruiz, Solar-Lezama, and Tang}{Singh et~al\mbox{.}}{2017}]%
        {singh2017generating}
\bibfield{author}{\bibinfo{person}{Rohit Singh}, \bibinfo{person}{Vamsi
  Meduri}, \bibinfo{person}{Ahmed Elmagarmid}, \bibinfo{person}{Samuel Madden},
  \bibinfo{person}{Paolo Papotti}, \bibinfo{person}{Jorge-Arnulfo
  Quian{\'e}-Ruiz}, \bibinfo{person}{Armando Solar-Lezama}, {and}
  \bibinfo{person}{Nan Tang}.} \bibinfo{year}{2017}\natexlab{}.
\newblock \showarticletitle{Generating concise entity matching rules}. In
  \bibinfo{booktitle}{\emph{Proceedings of the 2017 ACM International
  Conference on Management of Data}}. \bibinfo{pages}{1635--1638}.
\newblock


\bibitem[\protect\citeauthoryear{Steorts, Ventura, Sadinle, and
  Fienberg}{Steorts et~al\mbox{.}}{2014}]%
        {steorts2014comparison}
\bibfield{author}{\bibinfo{person}{Rebecca~C Steorts},
  \bibinfo{person}{Samuel~L Ventura}, \bibinfo{person}{Mauricio Sadinle}, {and}
  \bibinfo{person}{Stephen~E Fienberg}.} \bibinfo{year}{2014}\natexlab{}.
\newblock \showarticletitle{A comparison of blocking methods for record
  linkage}. In \bibinfo{booktitle}{\emph{International conference on privacy in
  statistical databases}}. Springer, \bibinfo{pages}{253--268}.
\newblock


\bibitem[\protect\citeauthoryear{Stonebraker, Bruckner, Ilyas, Beskales,
  Cherniack, Zdonik, Pagan, and Xu}{Stonebraker et~al\mbox{.}}{2013}]%
        {stonebraker2013data}
\bibfield{author}{\bibinfo{person}{Michael Stonebraker},
  \bibinfo{person}{Daniel Bruckner}, \bibinfo{person}{Ihab~F Ilyas},
  \bibinfo{person}{George Beskales}, \bibinfo{person}{Mitch Cherniack},
  \bibinfo{person}{Stanley~B Zdonik}, \bibinfo{person}{Alexander Pagan}, {and}
  \bibinfo{person}{Shan Xu}.} \bibinfo{year}{2013}\natexlab{}.
\newblock \showarticletitle{Data Curation at Scale: The Data Tamer System.}. In
  \bibinfo{booktitle}{\emph{Cidr}}, Vol.~\bibinfo{volume}{2013}.
\newblock


\bibitem[\protect\citeauthoryear{Suchanek, Kasneci, and Weikum}{Suchanek
  et~al\mbox{.}}{2007}]%
        {yago}
\bibfield{author}{\bibinfo{person}{Fabian~M. Suchanek},
  \bibinfo{person}{Gjergji Kasneci}, {and} \bibinfo{person}{Gerhard Weikum}.}
  \bibinfo{year}{2007}\natexlab{}.
\newblock \showarticletitle{Yago: A Core of Semantic Knowledge}. In
  \bibinfo{booktitle}{\emph{Proceedings of the 16th International Conference on
  World Wide Web}} (Banff, Alberta, Canada) \emph{(\bibinfo{series}{WWW '07})}.
  \bibinfo{pages}{697–706}.
\newblock


\bibitem[\protect\citeauthoryear{Vaswani, Shazeer, Parmar, Uszkoreit, Jones,
  Gomez, Kaiser, and Polosukhin}{Vaswani et~al\mbox{.}}{2017}]%
        {attention}
\bibfield{author}{\bibinfo{person}{Ashish Vaswani}, \bibinfo{person}{Noam
  Shazeer}, \bibinfo{person}{Niki Parmar}, \bibinfo{person}{Jakob Uszkoreit},
  \bibinfo{person}{Llion Jones}, \bibinfo{person}{Aidan~N. Gomez},
  \bibinfo{person}{\L{}ukasz Kaiser}, {and} \bibinfo{person}{Illia
  Polosukhin}.} \bibinfo{year}{2017}\natexlab{}.
\newblock \showarticletitle{Attention is All You Need}. In
  \bibinfo{booktitle}{\emph{Proceedings of the 31st International Conference on
  Neural Information Processing Systems}} (Long Beach, California, USA)
  \emph{(\bibinfo{series}{NIPS'17})}. \bibinfo{publisher}{Curran Associates
  Inc.}, \bibinfo{address}{Red Hook, NY, USA}, \bibinfo{pages}{6000–6010}.
\newblock
\showISBNx{9781510860964}


\bibitem[\protect\citeauthoryear{Vrande\v{c}i\'{c} and
  Kr\"{o}tzsch}{Vrande\v{c}i\'{c} and Kr\"{o}tzsch}{2014}]%
        {wikidata}
\bibfield{author}{\bibinfo{person}{Denny Vrande\v{c}i\'{c}} {and}
  \bibinfo{person}{Markus Kr\"{o}tzsch}.} \bibinfo{year}{2014}\natexlab{}.
\newblock \showarticletitle{Wikidata: A Free Collaborative Knowledgebase}.
\newblock \bibinfo{journal}{\emph{Commun. ACM}} \bibinfo{volume}{57},
  \bibinfo{number}{10} (\bibinfo{date}{Sept.} \bibinfo{year}{2014}),
  \bibinfo{pages}{78–85}.
\newblock
\showISSN{0001-0782}
\urldef\tempurl%
\url{https://doi.org/10.1145/2629489}
\showDOI{\tempurl}


\bibitem[\protect\citeauthoryear{Wang, Li, Yu, and Feng}{Wang
  et~al\mbox{.}}{2011}]%
        {wang2011entity}
\bibfield{author}{\bibinfo{person}{Jiannan Wang}, \bibinfo{person}{Guoliang
  Li}, \bibinfo{person}{Jeffrey~Xu Yu}, {and} \bibinfo{person}{Jianhua Feng}.}
  \bibinfo{year}{2011}\natexlab{}.
\newblock \showarticletitle{Entity matching: How similar is similar}.
\newblock \bibinfo{journal}{\emph{Proceedings of the VLDB Endowment}}
  \bibinfo{volume}{4}, \bibinfo{number}{10} (\bibinfo{year}{2011}),
  \bibinfo{pages}{622--633}.
\newblock


\bibitem[\protect\citeauthoryear{Wang, Li, Aslan, and Vinyals}{Wang
  et~al\mbox{.}}{2021}]%
        {wikigraphs}
\bibfield{author}{\bibinfo{person}{Luyu Wang}, \bibinfo{person}{Yujia Li},
  \bibinfo{person}{{\"{O}}zlem Aslan}, {and} \bibinfo{person}{Oriol Vinyals}.}
  \bibinfo{year}{2021}\natexlab{}.
\newblock \showarticletitle{WikiGraphs: {A} Wikipedia Text - Knowledge Graph
  Paired Dataset}.
\newblock \bibinfo{journal}{\emph{CoRR}}  \bibinfo{volume}{abs/2107.09556}
  (\bibinfo{year}{2021}).
\newblock
\showeprint[arXiv]{2107.09556}
\urldef\tempurl%
\url{https://arxiv.org/abs/2107.09556}
\showURL{%
\tempurl}


\bibitem[\protect\citeauthoryear{Wei and Zou}{Wei and Zou}{2019}]%
        {wei2019eda}
\bibfield{author}{\bibinfo{person}{Jason Wei} {and} \bibinfo{person}{Kai Zou}.}
  \bibinfo{year}{2019}\natexlab{}.
\newblock \showarticletitle{Eda: Easy data augmentation techniques for boosting
  performance on text classification tasks}.
\newblock \bibinfo{journal}{\emph{arXiv preprint arXiv:1901.11196}}
  (\bibinfo{year}{2019}).
\newblock


\bibitem[\protect\citeauthoryear{Weikum, Dong, Razniewski, and Suchanek}{Weikum
  et~al\mbox{.}}{2021}]%
        {kg_book}
\bibfield{author}{\bibinfo{person}{Gerhard Weikum}, \bibinfo{person}{Xin~Luna
  Dong}, \bibinfo{person}{Simon Razniewski}, {and} \bibinfo{person}{Fabian~M.
  Suchanek}.} \bibinfo{year}{2021}\natexlab{}.
\newblock \showarticletitle{Machine Knowledge: Creation and Curation of
  Comprehensive Knowledge Bases}.
\newblock \bibinfo{journal}{\emph{Found. Trends Databases}}
  \bibinfo{volume}{10}, \bibinfo{number}{2-4} (\bibinfo{year}{2021}),
  \bibinfo{pages}{108--490}.
\newblock
\urldef\tempurl%
\url{https://doi.org/10.1561/1900000064}
\showDOI{\tempurl}


\bibitem[\protect\citeauthoryear{Whang, Menestrina, Koutrika, Theobald, and
  Garcia-Molina}{Whang et~al\mbox{.}}{2009}]%
        {whang2009entity}
\bibfield{author}{\bibinfo{person}{Steven~Euijong Whang},
  \bibinfo{person}{David Menestrina}, \bibinfo{person}{Georgia Koutrika},
  \bibinfo{person}{Martin Theobald}, {and} \bibinfo{person}{Hector
  Garcia-Molina}.} \bibinfo{year}{2009}\natexlab{}.
\newblock \showarticletitle{Entity resolution with iterative blocking}. In
  \bibinfo{booktitle}{\emph{Proceedings of the 2009 ACM SIGMOD International
  Conference on Management of data}}. \bibinfo{pages}{219--232}.
\newblock


\bibitem[\protect\citeauthoryear{Wu, Zhang, Ilyas, and Rekatsinas}{Wu
  et~al\mbox{.}}{2020}]%
        {aimnet}
\bibfield{author}{\bibinfo{person}{Richard Wu}, \bibinfo{person}{Aoqian Zhang},
  \bibinfo{person}{Ihab Ilyas}, {and} \bibinfo{person}{Theodoros Rekatsinas}.}
  \bibinfo{year}{2020}\natexlab{}.
\newblock \showarticletitle{Attention-based Learning for Missing Data
  Imputation in HoloClean}. In \bibinfo{booktitle}{\emph{Proceedings of Machine
  Learning and Systems}}, Vol.~\bibinfo{volume}{2}. \bibinfo{pages}{307--325}.
\newblock


\bibitem[\protect\citeauthoryear{Yamada, Washio, Shindo, and Matsumoto}{Yamada
  et~al\mbox{.}}{2019}]%
        {Yamada2019GlobalED}
\bibfield{author}{\bibinfo{person}{Ikuya Yamada}, \bibinfo{person}{Koki
  Washio}, \bibinfo{person}{Hiroyuki Shindo}, {and} \bibinfo{person}{Yuji
  Matsumoto}.} \bibinfo{year}{2019}\natexlab{}.
\newblock \showarticletitle{Global Entity Disambiguation with Pretrained
  Contextualized Embeddings of Words and Entities}.
\newblock \bibinfo{journal}{\emph{arXiv: Computation and Language}}
  (\bibinfo{year}{2019}).
\newblock


\bibitem[\protect\citeauthoryear{Yan, Meyles, Haghighi, and Suciu}{Yan
  et~al\mbox{.}}{2020}]%
        {yan2020entity}
\bibfield{author}{\bibinfo{person}{Yan Yan}, \bibinfo{person}{Stephen Meyles},
  \bibinfo{person}{Aria Haghighi}, {and} \bibinfo{person}{Dan Suciu}.}
  \bibinfo{year}{2020}\natexlab{}.
\newblock \showarticletitle{Entity Matching in the Wild: A Consistent and
  Versatile Framework to Unify Data in Industrial Applications}. In
  \bibinfo{booktitle}{\emph{Proceedings of the 2020 ACM SIGMOD International
  Conference on Management of Data}}. \bibinfo{pages}{2287--2301}.
\newblock


\bibitem[\protect\citeauthoryear{Yang, Yih, He, Gao, and Deng}{Yang
  et~al\mbox{.}}{2015}]%
        {distmult}
\bibfield{author}{\bibinfo{person}{Bishan Yang}, \bibinfo{person}{Wen{-}tau
  Yih}, \bibinfo{person}{Xiaodong He}, \bibinfo{person}{Jianfeng Gao}, {and}
  \bibinfo{person}{Li Deng}.} \bibinfo{year}{2015}\natexlab{}.
\newblock \showarticletitle{Embedding Entities and Relations for Learning and
  Inference in Knowledge Bases}. In \bibinfo{booktitle}{\emph{3rd International
  Conference on Learning Representations, {ICLR} 2015, San Diego, CA, USA, May
  7-9, 2015, Conference Track Proceedings}}.
\newblock


\bibitem[\protect\citeauthoryear{Yang}{Yang}{2019}]%
        {yang2019aligraph}
\bibfield{author}{\bibinfo{person}{Hongxia Yang}.}
  \bibinfo{year}{2019}\natexlab{}.
\newblock \showarticletitle{Aligraph: A comprehensive graph neural network
  platform}. In \bibinfo{booktitle}{\emph{Proceedings of the 25th ACM SIGKDD
  international conference on knowledge discovery \& data mining}}.
  \bibinfo{pages}{3165--3166}.
\newblock


\bibitem[\protect\citeauthoryear{Zheng, Song, Ma, Tan, Ye, Dong, Xiong, Zhang,
  and Karypis}{Zheng et~al\mbox{.}}{2020}]%
        {dgl_ke}
\bibfield{author}{\bibinfo{person}{Da Zheng}, \bibinfo{person}{Xiang Song},
  \bibinfo{person}{Chao Ma}, \bibinfo{person}{Zeyuan Tan},
  \bibinfo{person}{Zihao Ye}, \bibinfo{person}{Jin Dong}, \bibinfo{person}{Hao
  Xiong}, \bibinfo{person}{Zheng Zhang}, {and} \bibinfo{person}{George
  Karypis}.} \bibinfo{year}{2020}\natexlab{}.
\newblock \showarticletitle{DGL-KE: Training Knowledge Graph Embeddings at
  Scale}. In \bibinfo{booktitle}{\emph{Proceedings of the 43rd International
  ACM SIGIR Conference on Research and Development in Information Retrieval}}
  \emph{(\bibinfo{series}{SIGIR '20})}. \bibinfo{publisher}{Association for
  Computing Machinery}, \bibinfo{address}{New York, NY, USA},
  \bibinfo{pages}{739–748}.
\newblock
\showISBNx{9781450380164}
\urldef\tempurl%
\url{https://doi.org/10.1145/3397271.3401172}
\showDOI{\tempurl}


\bibitem[\protect\citeauthoryear{Zhuge, Garcia-Molina, Hammer, and Widom}{Zhuge
  et~al\mbox{.}}{1995}]%
        {zhuge1995view}
\bibfield{author}{\bibinfo{person}{Yue Zhuge}, \bibinfo{person}{Hector
  Garcia-Molina}, \bibinfo{person}{Joachim Hammer}, {and}
  \bibinfo{person}{Jennifer Widom}.} \bibinfo{year}{1995}\natexlab{}.
\newblock \showarticletitle{View maintenance in a warehousing environment}. In
  \bibinfo{booktitle}{\emph{Proceedings of the 1995 ACM SIGMOD international
  conference on Management of data}}. \bibinfo{pages}{316--327}.
\newblock


\end{thebibliography}
